\documentclass[onecolumn,tightenlines,showpacs,nofootinbib,preprintnumbers,
amsmath,amssymb]{revtex4}
\usepackage[usenames,dvipsnames]{color}

\usepackage{graphicx}% Include figure files
\usepackage{bm}% bold math
\usepackage{lscape}

\begin{document}
%\nofiles

%\preprint{APS/123-QED}

\title{Realistic Earth matter effects and a method to acquire information about small $\theta_{13}$ in the detection of supernova neutrinos}
% Force line breaks with \\

\author{Xin-Heng Guo$^{1}$\footnote{Email: xhguo@bnu.edu.cn }, Ming-Yang Huang$^{1}$\footnote{Corresponding author email: hmy19151905@mail.bnu.edu.cn},
and Bing-Lin Young$^{2,3}$\footnote{Email: young@iastate.edu}}
\affiliation{\small $^{1}$College of Nuclear Science and
Technology, Beijing Normal University, Beijing 100875, China \\
\small $^2$ Department of Physics and Astronomy, Iowa State
University, Ames, Iowa 5001, USA \\
\small $^3$ Institute of Theoretical Physics, Chinese Academy of
Sciences, Beijing, China}

%\date{\today}

\begin{abstract}
In this paper, we first calculate the realistic Earth matter
effects in the detection of type II supernova neutrinos at the
Daya Bay reactor neutrino experiment which is currently under
construction. It is found that the Earth matter effects depend on
the neutrino incident angle $\theta$, the neutrino mass hierarchy
$\Delta m_{31}^{2}$, the crossing probability at the high
resonance region inside the supernova, $P_{H}$, the neutrino
temperature, $T_{\alpha}$, and the pinching parameter in the
neutrino spectrum, $\eta_{\alpha}$. We also take into account the
collective effects due to neutrino-neutrino interactions inside
the supernova. With the expression for the dependence of $P_H$ on
the neutrino mixing angle $\theta_{13}$, we obtain the relations
between $\theta_{13}$ and the event numbers for various reaction
channels of supernova neutrinos. Using these relations, we propose
a possible method to acquire information about $\theta_{13}$
smaller than $1.5^\circ$.
% by detecting the event numbers of supernova neutrinos.
Such a sensitivity cannot yet be achieved by the Daya Bay reactor
neutrino experiment which has a sensitivity of the order of
$\theta_{13}\sim 3^\circ$. Furthermore, we apply this method to
other neutrino experiments, i.e. Super-K, SNO, KamLAND, LVD,
MinBooNE, Borexino, and Double-Chooz. We also study the energy
spectra of the differential event numbers, ${\rm d }N/{\rm d }E$.
\end{abstract}

\maketitle

\section{\label{intro}Introduction}

Observable effects of supernova (SN) neutrinos in underground
detectors have been a subject of intense investigation in
astroparticle physics, both on general grounds and in relation to
the SN event like 1987A \cite{SN1987}. In particular, flavor
oscillation in SN may shed light on the problem of neutrino masses
and mixing by means of the associated matter effects. Several
neutrino laboratories, including the Daya Bay reactor neutrino
underground laboratory
 %%%
\cite{DayaBay} which is under construction, can be used to detect
possible neutrino events from an SN explosion and serve as the SN
Earth Warning System \cite{SNEWS}. Hence theoretical prediction
for the detection of SN neutrinos in the Daya Bay and other
neutrino experiments is very desirable.

In Ref. \cite{EM1}, the authors studied the matter effects,
including both the Earth and SN, on the detection of neutrinos
from a type II SN explosion. The results of the study were applied
to the Daya Bay reactor neutrino experiment which measures the
mixing angle $\theta_{13}$ down to $\sin^22\theta_{13}=0.01$. In
that paper, a simplified picture for the matter density of the
Earth was used, i.e. the mantle-core-mantle picture in which
$\rho=12g/{cm}^3$ ($\rho$ is the Earth matter density) for the
core and $\rho=5g/{cm}^3$ for the mantle, where the core radius
and the thickness of the mantle are half of the Earth radius,
respectively.  The energy spectra of neutrinos from the cooling
stage were assumed as exactly thermal and the average temperature
for the neutrinos, $T_{\nu_e}=3.5MeV$, $T_{\bar{\nu}_e}=5MeV$, and
$T_{\nu_x}=8MeV$, were used ($\nu_x$ denotes $\nu_{\mu}$,
$\bar{\nu}_{\mu}$, $\nu_{\tau}$, or $\bar{\nu}_{\tau}$). Due to
the large mixing angle solution of neutrinos \cite{EM2}, the
crossing probability at the low resonance region inside the SN,
$P_L$, is zero. Since the crossing probability at the high
resonance region, $P_H$, depends on the unknown neutrino mixing
angle $\theta_{13}$, two extreme cases, i.e. $P_{H}$ being $0$ and
$1$, were studied. In the present work, we improve the
investigation in Ref. \cite{EM1} by generalizing the simplified
mantle-core-mantle picture for the Earth matter density to the
realistic density profile including the effects due to variations
of the neutrino temperature $T_{\alpha}$ and the pinching
parameter $\eta_{\alpha}$ in the neutrino energy spectra. We also
include the collective effects arising from neutrino-neutrino
interactions in the SN. Furthermore, the target material is taken
to be the Linear Alkyl Benzene (LAB), which has been selected as
the main part of the liquid scintillator in the Daya Bay
experiment \cite{DayaBay}, instead of $C_9H_{12}$ as used in Ref.
\cite{EM1}.

In the realistic case, the matter density of the Earth changes
with the depth continuously \cite{Earth}, the energy spectra of
neutrinos from the cooling stage are not exactly thermal and has a
non-vanishing pinching parameter, and the neutrino temperature may
vary in some ranges \cite{EM2}\cite{Janka1}\cite{Janka2}.
Furthermore, since neutrino and anti-neutrino densities are very
high near the neutrinosphere, the neutrino-neutrino interarcions
become significance. This leads to collective effects in the SN
which affect the neutrino spectra\cite{Duan1}-\cite{Fogli}. One
main purpose of the present work is to investigate the effects of
all these factors on the detection of SN neutrinos. In order to
compare with Ref. \cite{EM1}, we will first calculate the effects
of the realistic distribution of the Earth matter density on the
detection of SN neutrinos in the simple case $T_{\nu_e}$ =
$3.5MeV$, $T_{\bar{\nu}_e}$ = $5MeV$, $T_x$ = $8MeV$, in the
absence of pinching parameters, i.e., $\eta_{\nu_e}$ =
$\eta_{\bar{\nu}_e}$ = $\eta_{\nu_x}$ = $0$. Then we study the
effects due to the variation of neutrino temperatures and the
inclusion of the pinching parameters in the energy spectra. In the
future, if the values of $T_{\alpha}$ and $\eta_{\alpha}$ can be
constrained more precisely, the event number of SN neutrinos can
be more accurately predicted.

The other main purpose of the present paper is to propose a
possible method to acquire information about $\theta_{13}$ below
the sensitivity of the Daya Bay experiment, i.e. $3^{\circ}$,
through the detection of SN neutrinos. This can be achieved since
at the high resonance region inside the SN, the crossing
probability $P_H$ is very sensitive to $\theta_{13}$.
%In the eightieth decade of last century, some authors gave the
Analytical expressions for transition probabilities between
different neutrino flavors have been investigated in the
literatures. Using the Landau-Zener formula \cite{Landau}, the
crossing probability, $P_C$, which is the neutrino jumping
probability from one mass eigenstate to another at the resonance
region, were also calculated for specific matter density
distributions, including the linear
\cite{MSW1}\cite{MSW2}\cite{MSW3}\cite{MSW4}, exponential
\cite{MSW5}, hyperbolic tangent \cite{MSW6}, and $1/r$
\cite{MSW7}\cite{MSW8} density distributions. Using a similar
method, the crossing probabilities at the high and low resonance
regions inside the SN, $P_H$ and $P_L$, were given in
\cite{MSW8}\cite{PH1}\cite{PH2}. The neutrino event numbers $N$
which we will obtain depend on $P_H$. Using the relation between
$P_H$ and $\theta_{13}$, we can predict the SN neutrino event
numbers $N$ as a function of $\theta_{13}$. With this, we give a
possible method to acquire information about $\theta_{13}$ smaller
than $1.5^{\circ}$ by measuring the ratios of the event numbers of
different flavor of SN neutrinos, e.g., the ratio of the event
numbers of $\nu_e$ and $\bar{\nu}_e$. This method is applied to
the existing neutrino experiments and experiments under
construction, including Daya Bay, Super-K, SNO, KamLAND, LVD,
MinBooNE, Borexino, and Double-Chooz.

Finally, the energy spectra of the differential neutrino event
numbers, ${\rm d }N/{\rm d }E$, where $E$ is the neutrino energy,
will be studied. We will make predictions for ${\rm d }N/{\rm d
}E$ as a function of $E$ for different neutrino Earth incident
angles $\theta$ and for different values of $\theta_{13}$. Then we
discuss some interesting properties of these predictions.

The paper is organized as follows. In Sec. II, we review the
necessary formulas for the detection of SN neutrinos on the Earth.
In Sec. III, we apply these formulas to the
%%%do numerical calculations for
Daya Bay experiment, where the realistic Earth matter density, the
collective effects, the variation of the neutrino temperature, and
the pinching of the neutrino spectra are considered.  In Sec. IV,
the relation between the neutrino event number $N$ and
$\theta_{13}$ is discussed and a method to obtain information of
about $\theta_{13}$ smaller than $1.5^\circ$ is proposed. This
method is applied to the Daya Bay experiment as an explicit
example. In Sec. V, the method is applied to several existing main
neutrino experiments, including Super-K, SNO, KamLAND, LVD,
MinBooNE, Borexino, and another experiment under construction
Double-Chooz. In Sec. VI, we study the energy spectra of the
differential neutrino event numbers (${dN}/{dE}$) and discuss some
of their properties. Finally, a summary is given in the concluding
section VII.

\section{\label{sec:hamckm}formulas for detection of SN neutrinos on the Earth}

 %%%
The SN explosion is one of the most spectacular cosmic events and
a source of new physical ideas. A broad area of topics of
fundamental physics can be studied by the observation of SN. In
the core collapse of SN, a vast amount of neutrinos are produced
in two bursts. In the first burst which lasts for only a few
milliseconds, electron neutrinos are generated via the inverse
beta-decay process which leads to a neutron rich star. In the
second burst which lasts longer, neutrinos of all flavors are
produced via the
 %%%
nucleon-nucleon bremsstrahlung process, $\nu_{\rm e}\bar\nu_{\rm
e}$ annihilation, and $e^+e^-$ annihilation
\cite{SNsum}\cite{Dighe}\cite{SNbook}.

When the SN neutrinos of each flavor are produced they are
approximately the effective mass eigenstates due to the extremely
high matter density environment. While they propagate outward to
the surface of the SN they could experience collective effects
arising from neutrino-neutrino interactions
\cite{Duan1}-\cite{Fogli} and the well-known
Mikheyev-Smirnov-Wolfenstein (MSW) effects
\cite{MSW1}-\cite{MSW8}. After travelling the cosmic distance to
reach the Earth, the arriving neutrinos are mass eigenstates,
which then oscillate in flavors while going through the Earth
matter. Therefore, we also have to consider the Earth matter
effects \cite{EM1}\cite{EM2}\cite{EM3}\cite{EM4}\cite{EM5} when we
compute the event numbers of the various flavors of neutrions.

In the Day Bay experiment, SN neutrinos undergo the following
reactions in the detector.

(1) $\bar{\nu}_{e}+p$ reaction

The large cross section, low threshold, and abundance of target
protons make this the dominant channel for the detection of SN
neutrinos. The inverse beta-decay process,
\begin{eqnarray}
\bar{\nu}_e+p\rightarrow e^{+}+n, \nonumber
\end{eqnarray}
has a reaction threshold \cite{Totani}\cite{Chen}
\begin{equation}
 E_{th}=1.80MeV. \label{Eth}
\end{equation}
At low energies we approximate the large cross section as
\cite{Chen}\cite{Burrows1}
\begin{equation}
 \sigma(\bar{\nu}_ep)=9.5\times10^{-44}(E(MeV)-1.29)^2 cm^2. \label{sigma p}
\end{equation}

(2) $\nu-e^{-}$ scattering

The neutrino-electron scattering,
\begin{eqnarray}
&&\ \nu_{e}+e^-\rightarrow \nu_{e}+e^- \quad({\rm CC\quad and\quad NC}), \nonumber\\
&&\ \bar{\nu}_{e}+e^-\rightarrow\bar{\nu}_{e}+e^-\quad({\rm CC\quad and\quad NC}),\nonumber\\
&&\ \nu_{\alpha}+e^-\rightarrow \nu_{\alpha}+e^- \quad({\rm NC}), \nonumber\\
&&\ \bar{\nu}_{\alpha}+e^-\rightarrow \bar{\nu}_{\alpha}+e^-
\quad({\rm NC}), \quad \alpha=\mu,\tau  \nonumber
\end{eqnarray}
where CC and NC stand respectively for the changed current and
neutral current interactions, produce recoil electrons with energy
from zero up to the kinematics maximum. In our rate calculation we
integrate over the ranges of the electron recoil energies. Then,
the total cross sections for the neutrino-electron scattering are
linearly proportional to the neutrino energy, and have the
following forms \cite{Chen}:
\begin{eqnarray}
 &&\ \sigma(\nu_ee\rightarrow\nu_ee)=9.20\times10^{-45}E(MeV)
 cm^2,\label{5} \nonumber \\
 &&\ \sigma(\bar{\nu}_ee\rightarrow\bar{\nu}_ee)=3.83\times10^{-45}E(MeV)
 cm^2, \nonumber \\
 &&\ \sigma(\nu_{\mu,\tau}e\rightarrow\nu_{\mu,\tau}e)=1.57\times10^{-45}E(MeV)
 cm^2, \nonumber \\
 &&\ \sigma(\bar{\nu}_{\mu,\tau}e\rightarrow\bar{\nu}_{\mu,\tau}e)=1.29\times10^{-45}E(MeV)
 cm^2. \label{sigma e}
\end{eqnarray}

(3) $^{12}C$ reactions

For the neutrinos and $^{12}C$ system, there are two
charged-current and six neutral-current reactions. The effective
cross sections are obtained by scaling the experimentally measured
energy values from the decay of the muon at rest to the energy
scale for SN neutrinos. In this way, one can obtain the following
effective cross sections
\cite{Chen}\cite{Burrows1}\cite{Burrows2}:

Charged-current capture of $\Bar{\nu}_e$:
\begin{eqnarray}
 &&\ \Bar{\nu}_e+^{12}C\rightarrow^{12}B+e^+, \quad E_{th}=14.39MeV,
\nonumber \\
 &&\ ^{12}B\rightarrow^{12}C+e^{-}+\Bar{\nu}_e, \nonumber \\
 &&\ \langle\sigma(^{12}C(\bar{\nu}_e,e^{+})^{12}B)\rangle=1.87\times10^{-42}
cm^2.
 \label{sigma CB}
\end{eqnarray}

Charged-current capture of $\nu_e$:
\begin{eqnarray}
&&\ \nu_e+^{12}C\rightarrow^{12}N+e^-, \quad E_{th}=17.34MeV,
\nonumber
\\
&&\ ^{12}N\rightarrow^{12}C+e^{+}+\nu_e,
\nonumber \\
&&\
\langle\sigma(^{12}C(\nu_e,e^{-})^{12}N)\rangle=1.85\times10^{-43}
cm^2. \label{sigma CN}
\end{eqnarray}

Neutral-current inelastic scattering of $\nu_{\alpha}$ or
$\Bar{\nu}_{\alpha}$ where $\alpha=e, \mu, \tau$:
\begin{eqnarray}
&&\ \nu_{\alpha}+^{12}C\rightarrow^{12}C^{\ast}+\nu_{\alpha}^{'},
\quad E_{th}=15.11MeV,
\nonumber \\
&&\
\bar{\nu}_{\alpha}+^{12}C\rightarrow^{12}C^{\ast}+\bar{\nu}_{\alpha}^{'},
\quad E_{th}=15.11MeV, \nonumber\\
 &&\ ^{12}C^{\ast}\rightarrow^{12}C+\gamma,
\nonumber \\
 &&\ \langle\sigma(\nu_e^{12}C)\rangle=1.33\times10^{-43}cm^2, \nonumber \\
 &&\ \langle\sigma(\Bar{\nu}_e^{12}C)\rangle=6.88\times10^{-43}cm^2, \nonumber \\
 &&\
 \langle\sigma(\nu_x(\Bar{\nu}_x)^{12}C)\rangle=3.73\times10^{-42}cm^2, \quad x=\mu,\tau.
 \label{sigma CC}
\end{eqnarray}

The effective cross sections in Eqs.
%(\ref{sigma CB})(\ref{sigma CN})(\ref{sigma CC})
(\ref{sigma CB})-(\ref{sigma CC}) are given for SN neutrinos
without oscillations \cite{EM1}\cite{Chen}. When neutrino
oscillations are taken into account, the oscillations of higher
energy $\nu_x$ into $\nu_e$ result in an increased event rate
since the expected $\nu_e$ energies are just at or below the
charged-current reaction threshold. This leads to an increase by a
factor of 35 for the cross section
$\langle\sigma(^{12}C(\nu_e,e^-)^{12}N)\rangle$ if we average it
over a $\nu_e$ distribution with $T=8MeV$ rather than $3.5MeV$.
Similarly, the cross section
$\langle\sigma(^{12}C(\Bar{\nu}_e,e^+)^{12}B)\rangle$ is increased
by a factor of 5. For the case of neutral-current inelastic
scattering of $\nu_{\alpha}$ or $\Bar{\nu}_{\alpha}$, when the
oscillations of higher energy $\nu_x$ into $\nu_e$ and $\bar\nu_x$
into $\bar\nu_e$ are taken into account, the cross section
$\langle\sigma(\nu_e^{12}C)\rangle$ is increased by a factor of 28
%while when the oscillation of higher energy $\nu_x$ into
%$\Bar{\nu}_e$ is considered,
and the cross section $\langle\sigma(\Bar{\nu}_e^{12}C)\rangle$ is
increased by a factor of 5.

There will be several detectors located at the near and far sites
at the Daya Bay experiment. In Ref. \cite{EM1}, the authors take
the liquid scintillator to be mainly $C_9H_{12}$ and the total
detector mass to be 300 tons. The Daya Bay Collaboration has
decided to use LAB as the main part of the liquid scintillator and
the total detector mass is about 300 tons. LAB, which has a
chemical composition including $C$ and $H$, is a mixture of the
mono-alkyl benzene with 9 to 14 carbon atoms in the side chain.
The main LAB components contain 10 to 13 carbon atoms in the side
chain. Therefore, LAB can be approximately expressed as
$C_6H_5-C_nH_{2n+1}$ where $n=9\sim14$. In our calculation, the
ratio of the numbers of $C$ and $H$, $N_C/N_H$, is about 0.6. Then
the total numbers of target protons, electrons, and $^{12}C$ are
\begin{eqnarray}
N_T^{(p)}=2.20\times10^{31},  \nonumber\\
N_T^{(e)}=1.01\times10^{32},  \nonumber\\
N_T^{(C)}=1.32\times10^{31} \label{NC}.
\end{eqnarray}
%respectively.

Neutrinos from an SN may travel through a significant portion of
the Earth before reaching the detector and are therefore subject
to the Earth matter effects. Suppose a neutrino reaches the
detector with the incident angle $\theta$ as indicated in Fig. 1,
then the distance the neutrino travelling through the Earth is
\begin{equation}
 L=(-R+h)\cos{\theta}+\sqrt{R^2-(R-h)^2\sin^2{\theta}}, \label{L}
\end{equation}
where $h$ is the underground depth of the detector and $R$
($6400km$) is the radius of the Earth. At the Daya Bay experiment,
$h\approx0.4km$. Let $x$ be the distance that the neutrino travels
into the Earth, then the distance of the neutrino to the center of
the Earth, $\tilde{x}$, is given by
\begin{equation}
 \tilde{x}=\sqrt{(-R+h)^2+(L-x)^2+2(R-h)(L-x)\cos{\theta}}. \label{x}
\end{equation}

In the following, we will calculate the event numbers of SN
neutrinos that can be observed through various reaction channels
 %%%
"$i$" at the Daya Bay experiment. This will be done by integrating
over the neutrino energy $E$, the product of the target number
$N_T$, the cross section of each channel $\sigma$, and the
neutrino flux function $F_{\alpha}^D(E)/4\pi D^2$,
\begin{equation}
 N_\alpha (i)=N_T\int{{\rm d}E\cdot\sigma(i)\cdot\frac{1}{4\pi
 D^2}\cdot F_{\alpha}^D}, \label{Ntotal}
\end{equation}
where $\alpha$ stands for the neutrino or antineutrino of a given
flavor, $D$ is the distance between the SN and the Earth, and the
index $i$ represents different channels through which SN neutrinos
are observed.

For the neutrino of flavor $\alpha$, the time-integrated neutrino
energy spectra can be described by the Fermi-Dirac distribution
(we consider the case where the spectra of neutrinos from the
cooling stage are not exactly thermal)
\cite{EM2}\cite{Janka1}\cite{Janka2}\cite{SNsum}\cite{Engel},
\begin{equation}
 F_{\alpha}^{(0)}(E)=\frac{N_{\alpha}^{(0)}}{F_{\alpha
 2}T_{\alpha}^{3}}\frac{E^2}{\exp{(E/T_{\alpha}-\eta_\alpha)}+1},
 \label{F0a1}
\end{equation}
where $T_{\alpha}$ is the typical temperature of the neutrino
\cite{EM2}\cite{Janka1}\cite{Janka2},
\begin{equation}
 T_{\nu_e}=3-4MeV,\quad T_{\bar{\nu}_e}=5-6MeV,\quad
 T_{\nu_x}=7-9MeV, \label{T}
 \quad (\nu_x=\nu_{\mu},\nu_{\tau},\bar{\nu}_{\mu},\bar{\nu}_{\tau}),
\end{equation}
and $\eta_\alpha$ is the pinching parameter of the spectra
($\eta_\alpha >0$) to represent the deviation from being exactly
thermal. The values of $\eta_\alpha$ for $\nu_x$ and $\bar{\nu}_x$
($x=\mu,\tau$) are the same (which will be denoted as
$\eta_{\nu_x}$ in the following) since they have the same
interactions, and are in general different from $\eta_{\nu_e}$ or
$\eta_{\bar{\nu}_e}$. The values of $\eta_\alpha$ need not be
constant throughout the cooling stage, and are typically
\cite{EM2}\cite{Janka1}\cite{Janka2}
\begin{equation}
 \eta_{\nu_e}\approx3-5, \quad \eta_{\bar{\nu}_e}\approx2.0-2.5,\quad
 \eta_{\nu_x}\approx0-2, \label{eta}
  \quad
  (\nu_x=\nu_{\mu},\nu_{\tau},\bar{\nu}_{\mu},\bar{\nu}_{\tau}).
\end{equation}
In Eq. (\ref{F0a1}), $F_{\alpha j}$, where $j$ is an integer, is
defined by
\begin{equation}
 F_{\alpha j}=\int_{0}^{\infty}\frac{x^j}{\exp{(x-\eta_\alpha)}+1}{\rm
 d}x, \label{Fa}
\end{equation}
and $N_\alpha^{(0)}$ is the total number of the neutrinos of
flavor $\alpha$,
\begin{equation}
 N_\alpha^{(0)}=\frac{L_\alpha^{(0)}}{\langle E_\alpha^{(0)}\rangle},
 \label{N0}
\end{equation}
where the average neutrino energy is $\langle
E_\alpha^{(0)}\rangle=\frac{F_{\alpha 3}}{F_{\alpha_2}}T_\alpha$
and the luminosity $L_\alpha^{(0)}$ is related to the total energy
release during the SN explosion, $E_{SN}^{(0)}$, through the
following equation:
\begin{equation}
L_\alpha^{(0)}=\frac{0.99}{6}E_{SN}^{(0)}. \label{L0}
\end{equation}
In the numerical calculations
 %%%
below we take
\begin{equation}
E_{SN}^{(0)}=1.97\times10^{59}MeV, \label{E0}
\end{equation}
and the distance $D$ to be 10 kpc =$3.09\times10^{22}cm$
\cite{Bahcall}\cite{Ahrens}. Using the above formulas, the energy
spectrum function can be rewritten as
\begin{equation}
 F_{\alpha}^{(0)}(E)=\frac{L_\alpha^{(0)}}{F_{\alpha
 3}T_{\alpha}^{4}}\frac{E^2}{\exp{(E/T_{\alpha}-\eta_\alpha)}+1}.
 \label{Foa2}
\end{equation}

In order to obtain the neutrino energy spectrum function at the
detector, the collective effects \cite{Duan1}-\cite{Fogli}, the
MSW effects \cite{MSW1}-\cite{MSW8}, and the Earth matter effects
\cite{EM2}\cite{EM3}\cite{EM4}\cite{EM5} should be considered. Let
$P_H$ ($P_L$) be the crossing probability at the high (low)
resonance regions inside the SN, $P_{\nu\nu}$ represent the
collective effects of neutrino-neutrino interactions which is a
stepwise flavor conversion probability of neutrino at a critical
energy $E_C$, and $P_{ie}$ ($i=1,2,3$) be the probability that a
neutrino mass eigenstate $\nu_i$ enters the surface of the Earth
and arrives at the detector as an electron neutrino $\nu_e$. Then
the flux of $\nu_e$ at the detector, denoted as $F_{\nu_e}^D$, can
be written as
\begin{equation}
F_{\nu_e}^D=\sum_i P_{ie}F_i, \label{FD}
\end{equation}
where $F_i$ is the flux of $\nu_i$ at the Earth surface, in either
the normal or inverted hierarchy. $P_{ie}$ is the probability of
the $i$th mass eigenstate contained in $\nu_e$ and obeys the
unitary condition $\sum_i P_{ie}=1$.

A significant amount of studies on the collective effects of
neutrino-neutrino interactions, including simulations, have been
made by a number of authors, e.g., Duan {\it et al.}
\cite{Duan1}\cite{Duan2}\cite{Duan3}, Dusgupta {\it et al.}
\cite{Dasgupta1}, Raffelt {\it et al.}
\cite{Raffelt1}\cite{Raffelt2}, Esteban-Pretel {\it et al.}
\cite{Esteban-Pretel}, and Fogli {\it et al.} \cite{Fogli}. In
order to obtain a simple expression of the stepwise flavor
conversion probabilities $P_{\nu\nu}$ for the neutrino and
$\bar{P}_{\nu\nu}$ for the antineutrino, we take a constant matter
density and box-spectra for both the neutrino and antineutrino
\cite{Raffelt2}. An analysis of the collective effects in the case
of three flavors has been made in \cite{Dasgupta1}, and it allows
us to characterize the collective oscillation effects and to write
down the flavor spectra of the neutrino and antineutrino arriving
at the Earth. Following \cite{Dasgupta1}, we have
$P_{\nu\nu}=\bar{P}_{\nu\nu}=1$ in the case of normal hierarchy;
and $\bar{P}_{\nu\nu}=1$, while
\begin{equation}
P_{\nu\nu}=
\begin{cases} 1 & (E<E_C), \\
 0 & (E>E_C),
 \end{cases}
\label{Pnunu}
\end{equation}
in the case of inverted hierarchy, where $E_C=7MeV$
\cite{Dasgupta1}\cite{Fogli}.

Due to the large mixing angle solution of the neutrino mixing, the
crossing probability at the low resonance region inside the SN
vanishes, $P_L=\bar{P}_L=0$. Also, for very small
$\sin\theta_{13}$, we can neglect the contributions from $P_{3e}$
and $\bar{P}_{3e}$ \cite{EM3}. Therefore, after a straightforward
calculation the following results \cite{Dasgupta1} for the fluxes
at the detector can be obtained:
\begin{eqnarray}
F_{\nu_e}^{D(N)}&=&P_{2e}P_HF_{\nu_e}^{(0)}+(1-P_{2e}P_H)F_{\nu_x}^{(0)},
\nonumber \\
F_{\bar{\nu}_e}^{D(N)}&=&(1-\bar{P}_{2e})F_{\bar{\nu}_e}^{(0)}+\bar{P}_{2e}F_{\bar{\nu}_x}^{(0)},
\nonumber \\
2F_{\nu_x}^{D(N)}&=&(1-P_{2e}P_H)F_{\nu_e}^{(0)}+(1+P_{2e}P_H)F_{\nu_x}^{(0)},
\nonumber \\
2F_{\bar{\nu}_x}^{D(N)}&=&\bar{P}_{2e}F_{\bar{\nu}_e}^{(0)}+(2-\bar{P}_{2e})F_{\bar{\nu}_x}^{(0)},
\label{FDN}
\end{eqnarray}
for the normal hierarchy ($\triangle m_{31}^2>0$), and
\begin{eqnarray}
F_{\nu_e}^{D(I)}&=&
\begin{cases} P_{2e}F_{\nu_e}^{(0)}+(1-P_{2e})F_{\nu_x}^{(0)}, & (E<E_C) \\
 F_{\nu_x}^{(0)}, & (E>E_C)
 \end{cases}
\nonumber \\
F_{\bar{\nu}_e}^{D(I)}&=&\bar{P}_H(1-\bar{P}_{2e})F_{\bar{\nu}_e}^{(0)}+(1+\bar{P}_{2e}\bar{P}_H-\bar{P}_H)F_{\bar{\nu}_x}^{(0)},
\nonumber \\
2F_{\nu_x}^{D(I)}&=&
\begin{cases} (1-P_{2e})F_{\nu_e}^{(0)}+(1+P_{2e})F_{\nu_x}^{(0)}, & (E<E_C) \\
 F_{\nu_e}^{(0)}+F_{\nu_x}^{(0)}, & (E>E_C)
 \end{cases}
\nonumber \\
2F_{\bar{\nu}_x}^{D(I)}&=&(1+\bar{P}_{2e}\bar{P}_H-\bar{P}_H)F_{\bar{\nu}_e}^{(0)}+(1+\bar{P}_H-\bar{P}_{2e}\bar{P}_H)F_{\bar{\nu}_x}^{(0)},
\label{FDI}
\end{eqnarray}
for the inverted hierarchy ($\triangle m_{31}^2<0$).

Let us remark the important result that an unit flavor conversion
probability means the absence of the collective effects. Hence as
indicated in Eq. (\ref{FDN}) the final SN neutrino and
antineutrino fluxes in the normal hierarchy are not modified by
the collective effects. Neither are the antineutrino fluxes in the
inverted hierarchy. Only the neutrino fluxes in the inverted
hierarchy and for $E>E_C$ are modified as indicated in Eq.
(\ref{FDI}).

%In Eqs. (\ref{FDN}) and (\ref{FDI}),
The probability $P_{ie}$ ($i=1,2,3$) has been calculated in Ref.
\cite{EM3}, in particular,
\begin{equation}
P_{2e}=\sin^2\theta_{12}+\frac{1}{2}\sin^22{\theta_{12}}\int_{x_0}^{x_f}dxV(x)\sin\phi
_{x\rightarrow x_f}^{m}, \label{P2e}
\end{equation}
where $\theta_{12}=32.5^{\circ}$ \cite{Fogli2}, $V(x)$ is the
potential that the neutrino experiences in the Earth, and
$\phi_{a\rightarrow b}^m$ is defined as
\begin{equation}
\phi_{a\rightarrow b}^m=\int_a^b {\rm d}x \triangle_m(x),
\label{phi}
\end{equation}
where
\begin{equation}
\triangle_m(x)=\frac{\triangle
 m_{21}^2}{2E}\sqrt{(\cos2\theta_{12}-\varepsilon(x))^2+\sin^22\theta_{12}}.
 \label{Deta}
\end{equation}
In Eq. (\ref{Deta}), $\triangle m_{21}^2=7.1\times10^{-5} eV^2$
and $\varepsilon(x)$ is defined as
\begin{equation}
\varepsilon(x)=\frac{2EV(x)}{\triangle m_{21}^2}. \label{sigima}
\end{equation}
For a typical neutrino energy $E=10MeV$, $\varepsilon$ is less
than 0.13 \cite{EM1}. Therefore, we neglect
 %%%
contributions of $O(\varepsilon^2)$.

In the Earth the potential $V(x)$ is $\sqrt{2}G_FN_e(x)$ for an
electron neutrino and $-\sqrt{2}G_FN_e(x)$ for an electron
anti-neutrino where $G_F$ is the Fermi constant and $N_e(x)$ is
the electron number density in the Earth matter. Let the matter
mass density inside the Earth be $\rho(x)$. For nuclei of equal
number of protons and neutrons the electron number density is
\begin{equation}
N_e(x)=\rho(x)/(m_p+m_n), \label{Nep}
\end{equation}
where $m_p$ and $m_n$ are respectively the proton and neutron
masses. The realistic matter density inside the Earth is shown in
Table I \cite{Earth}.

Since the crossing probability at the high resonance region $P_H$
depends on the neutrino mixing-angle $\theta_{13}$, which is
unknown \cite{SNsum}\cite{Fogli2}\cite{Fogli3}, we will first
consider two extreme cases, $P_H=0$ (for the pure adiabatic
conversion) and $P_H=1$ (corresponding to a strong violation of
the adiabatic condition). This enables us to estimate the range of
the total neutrino event numbers to be given in Sec. III.  Then in
Sec. IV, we will obtain the expression of the crossing probability
$P_H$ and work out the relation between the event number $N$ and
$\theta_{13}$. From appropriate ratios of event number which
change with $\theta_{13}$, we will be able to obtain some
information about $\theta_{13}$ smaller than $1.5^\circ$ by
detecting SN neutrinos.

\section{\label{sec:cpv1}Numerical results for Earth matter effects
                        in the detection of SN neutrinos}
In this section, we calculate the realistic Earth matter effects
in the detection of type II SN neutrinos at the Daya Bay
experiment and give the numerical results for the two extreme
cases of $P_H= 1$ and 0.  A summary of the results is given in
Table II.

\subsection{\label{subsec:form}A simple case $T_{\nu_e}$ = $3.5MeV$,
$T_{\bar{\nu}_e}$ = $5MeV$, $T_x$ = $8MeV$, $\eta_{\nu_e}$ =
$\eta_{\bar{\nu}_e}$ = $\eta_{\nu_x}$ = $0$}

We consider, similar to Refs. \cite{EM1}\cite{Chen}, the simple
case of $T_{\nu_e}=3.5MeV$, $T_{\bar{\nu}_e}=5MeV$,
$T_{\nu_x}=8MeV$ ($\nu_x$ denotes $\nu_{\mu}$, $\bar{\nu}_{\mu}$,
$\nu_{\tau}$, or $\bar{\nu}_{\tau}$) and the neutrino spectra from
the cooling stage to be exactly thermal, i.e.,
$\eta_{\nu_e}=\eta_{\bar{\nu}_e}=\eta_{\nu_x}=0$. The realistic
Earth matter density we use is given in Table I \cite{Earth}. The
numerical results for the event numbers for various reaction
channels are given in Fig. 2.

The inverse beta-decay $\bar{\nu}_e+p\rightarrow e^++n$ has the
largest Earth matter effect among all the channels. It can be seen
from Fig. 2(a) that the maximum Earth matter effect appears at
around $\theta\sim94^\circ$ and the effect is about $6.82\%$ in
the cases of $P_H$ = 0, 1 (normal hierarchy) and $P_H=1$ (inverted
hierarchy). When the neutrino incident angle becomes larger than
about $103^\circ$, the event number increases very slowly. In the
case $P_H=0$ (inverted hierarchy), the event number is independent
of the incident angle.

For the neutrino-electron elastic scattering, we can see from Fig.
2(b) that the total event number is much smaller than that of the
inverse beta-decay. The maximum Earth matter effect appears at
$\theta\sim93^\circ$ and the amount is as large as $2.03\%$ for
$P_H=1$ (normal hierarchy), $0.36\%$ for $P_H=1$ (inverted
hierarchy), $0.36\%$ for $P_H=0$ (normal hierarchy), while there
is no Earth matter in the case $P_H=0$ (inverted hierarchy). When
the incident angle becomes larger than about $100^\circ$, the
total event numbers increase very slowly for all the above four
cases.
 %%%
The smallness of the event number of this channel is due to the
small cross sections of the neutrino-electron elastic scattering
given in Eq. (\ref{sigma e}).

For the neutrino-carbon scattering, it can be seen from Fig. 2(c)
that the maximum Earth matter effect appears at
$\theta\sim93^\circ$ and the amount is as large as $1.97\%$ for
$P_H=1$ (normal hierarchy), $0.99\%$ for $P_H=1$ (inverted
hierarchy), $0.99\%$ for $P_H=0$ (normal hierarchy), while there
is no Earth matter in the case $P_H=0$ (inverted hierarchy). When
the incident angle becomes larger than about $100^\circ$, the
total event numbers increase slowly for all the four cases.

\subsection{\label{subsec:form}Realistic Earth matter effects for
different $T_{\alpha}$ and $\eta_{\alpha}$}

In this subsection, we consider the realistic Earth matter effects
due to the variations of both the neutrino temperatures
$T_{\alpha}$
%in the ranges in Eq. (\ref{T})
and the pinching parameters $\eta_{\alpha}$.
%in the ranges in Eq. (\ref{eta}).
In the following, we will calculate the SN neutrino event numbers
for the limiting walue of $T_{\alpha}$ and $\eta_{\alpha}$ given
in Eqs. (\ref{T}) and (\ref{eta}). In other words, we consider the
event number of SN neutrinos in the following two extreme cases:

$\bullet$ $T_{\nu_e}=4MeV$, $T_{\bar{\nu}_e}=6MeV$,
$T_{\nu_x}=9MeV$, $\eta_{\nu_e}=5$, $\eta_{\bar{\nu}_e}=2.5$,
$\eta_{\nu_x}=2$;

$\bullet$ $T_{\nu_e}=3MeV$, $T_{\bar{\nu}_e}=5MeV$,
$T_{\nu_x}=7MeV$, $\eta_{\nu_e}=3$, $\eta_{\bar{\nu}_e}=2$,
$\eta_{\nu_x}=0$.

The results are shown in Figs. 3-4. \vspace{0.3cm}

(1)  The case $T_{\nu_e}=4MeV$, $T_{\bar{\nu}_e}=6MeV$,
$T_{\nu_x}=9MeV$, $\eta_{\nu_e}=5$, $\eta_{\bar{\nu}_e}=2.5$,
$\eta_{\nu_x}=2$ \vspace{0.3cm}

For the inverse beta-decay $\bar{\nu}_e+p\rightarrow e^++n$, it
can be seen from Fig. 3(a) that the maximum Earth matter effect
appears at around $\theta\sim95^\circ$ and the effect is about
$6.98\%$ in the cases where $P_H$ = 0, 1 (normal hierarchy) and
$P_H=1$ (inverted hierarchy). When the incident angle becomes
larger than about $103^\circ$, the event number increases very
slowly. In the case $P_H=0$ (inverted hierarchy), the event number
is independent of the incident angle.

We can see from Fig. 3(b) that the total event number of the
neutrino-electron elastic scattering is much smaller than that of
the inverse beta-decay. The maximum Earth matter effect appears at
$\theta\sim94^\circ$ and the amount is as large as $2.48\%$ for
$P_H=1$ (normal hierarchy), $0.43\%$ for $P_H=1$ (inverted
hierarchy), $0.43\%$ for $P_H=0$ (normal hierarchy), while there
is no Earth matter in the case $P_H=0$ (inverted hierarchy). When
the incident angle becomes larger than about $100^\circ$, the
total event numbers increase very slowly for all the above four
cases.

For the neutrino-carbon scattering, it can be seen from Fig. 3(c)
that the maximum Earth matter effect appears at
$\theta\sim93^\circ$ and the amount is as large as $2.86\%$ for
$P_H=1$ (normal hierarchy), $1.43\%$ for $P_H=1$ (inverted
hierarchy), $1.43\%$ for $P_H=0$ (normal hierarchy), while there
is no Earth matter in the case $P_H=0$ (inverted hierarchy). When
the incident angle becomes larger than about $100^\circ$, the
total event numbers increase slowly for all the four cases.
\vspace{0.3cm}

(2) The case $T_{\nu_e}=3MeV$, $T_{\bar{\nu}_e}=5MeV$,
$T_{\nu_x}=7MeV$, $\eta_{\nu_e}=3$, $\eta_{\bar{\nu}_e}=2$,
$\eta_{\nu_x}=0$ \vspace{0.3cm}

Similar to the above case, for the inverse beta-decay
$\bar{\nu}_e+p\rightarrow e^++n$, it can be seen from Fig. 4(a)
that the maximum Earth matter effect appears at around
$\theta\sim94^\circ$ and the effect is about $3.69\%$ in the cases
where $P_H$ = 0, 1 (normal hierarchy) and $P_H=1$ (inverted
hierarchy). When the incident angle becomes larger than about
$103^\circ$, the event number increases very slowly. In the case
$P_H=0$ (inverted hierarchy), the event number is independent of
the incident angle.

For the neutrino-electron elastic scattering, which is plotted in
Fig. 4(b), the maximum Earth matter effect appears at
$\theta\sim93^\circ$ and the amount is as large as $1.59\%$ for
$P_H=1$ (normal hierarchy), $0.21\%$ for $P_H=1$ (inverted
hierarchy), $0.21\%$ for $P_H=0$ (normal hierarchy), while there
is no Earth matter in the case $P_H=0$ (inverted hierarchy). When
the incident angle becomes larger than about $100^\circ$, the
total event numbers increase very slowly for all the above four
cases.

Also, for the neutrino-carbon scattering, it can be seen from
Fig.4(c) that the maximum Earth matter effect appears at
$\theta\sim93^\circ$ and the amount is as large as $1.62\%$ for
$P_H=1$ (normal hierarchy), $\theta\sim92^\circ$ and $0.81\%$ for
$P_H=1$ (inverted hierarchy), $\theta\sim92^\circ$ and $0.80\%$
for $P_H=0$ (normal hierarchy), while there is no Earth matter in
the case $P_H=0$ (inverted hierarchy). When the incident angle
becomes larger than about $100^\circ$, the total event numbers
increase slowly for all the four cases.

The above results can be understood by considering the fact that
the oscillation behavior is determined by the factor $\triangle
m_{21}^2(eV^2)L(m)/E(MeV)$, where $L$ is the distance that the
neutrino travels in the Earth as given in Eq. (\ref{L}).  When
$\theta<90^\circ$ this distance is smaller than $10~km$ and hence
the amount of the Earth matter effects are very small. When
$\theta$ increases beyond $90^\circ$ this distance exceeds
$100~km$, the Earth matter effect becomes greater and reaches a
maximum value for $91^\circ\sim95^\circ$. When $\theta$ is more
than $100^\circ$, the distance that neutrino travels in the Earth
is greater than $2000~km$, then there could be many oscillations
in $F_{\nu_e}^D$ and hence the averaging Earth matter effect is
smaller than the maximum value.

In Table II, we list the incipient values of the SN neutrino event
numbers where the neutrino incident angle $\theta$ is zero and
hence practically the vacuum event numbers, the minimum SN
neutrino event numbers and hence the maximum Earth matter effects,
the incident angles at which the maximum Earth matter effects
appear, and the Earth matter effects for all the cases considered.
It can be seen that the realistic Earth matter effects are much
smaller than those given in Ref. \cite{EM1} where the
mantle-core-mantle approximation for the Earth matter density was
used. When both $T_{\alpha}$ and $\eta_{\alpha}$ vary in their
ranges, the event numbers of SN neutrinos and the Earth matter
effects will vary all the three reactions. More precise values of
$T_{\alpha}$ and $\eta_{\alpha}$ will help obtain more reliable
event numbers and the Earth matter effects. There are also some
characters  which can be seen from Table II:

$\bullet$ In the two cases: $P_H=1$ (inverted hierarchy) and
$P_H=0$ (normal hierarchy), we have the same event numbers and the
same Earth matter effects for all the three kinds of reactions;

$\bullet$ In the case $P_H=0$ (inverted hierarchy), the event
numbers are independent of the incident angle $\theta$ (i.e. there
is no Earth matter effects) for all the three kinds of reactions;

$\bullet$ In the case $P_H=1$ (normal hierarchies), there are the
largest Earth matter effects in all the four cases for the three
kinds of reactions;

$\bullet$ When $T_\alpha$ and $\eta_\alpha$ increase, the maximum
Earth matter effects in all the four cases increase for the three
kinds of reactions.

\section{\label{sec:hamckm}A possible method to acquire information about $\theta_{13}$ smaller than $1.5^\circ$}

In the above section, we took the crossing probability at the high
resonance region inside the SN, $P_H$, to be in the two extreme
cases, i.e. $P_H = 0$ and $P_H = 1$, due to the fact that $P_H$
depends on the neutrino mixing angle $\theta_{13}$ which is
unknown. In this section, using the expression of $P_H$, we will
derive relations between the event numbers of SN neutrinos, $N$,
and $\theta_{13}$, and then propose a possible method to obtain
information about $\theta_{13}$ smaller than $1.5^\circ$.

%In the eightieth decade of last century,
Considering the MSW effects \cite{MSW1}, a number of authors
including Bethe \cite{MSW2}, Parke \cite{MSW3}, Haxton
\cite{MSW4}, Petcov {\it et al.} \cite{MSW5}, N\"{o}tzold
\cite{MSW6}, Kuo and Pantaleone \cite{MSW7}\cite{MSW8} used the
Landau-Zener formula \cite{Landau} to calculate the crossing
probability $P_C$ at the resonance region inside a star. Several
different density distributions in the case of two-flavor
transitions were considered. In particular in
\cite{MSW7}\cite{MSW8} the following result is given:
\begin{equation}
P_C=\frac{\exp(-\frac{\pi}{2}\gamma
F)-\exp[-\frac{\pi}{2}\gamma(\frac{F}{\sin^2\beta})]}{1-\exp[-\frac{\pi}{2}\gamma(\frac{F}{\sin^2\beta})]}
\label{GLPC},
\end{equation}
where $\beta$ is the mixing angle and
\begin{eqnarray}
 \gamma=\frac{|\Delta m^2|
}{2E}\frac{\sin^22\beta}{\cos2\beta}\frac{1}{|d\ln N_e/dr|_{res}},
\label{rngamma}
\end{eqnarray}
where $N_e$ is the electron density and $F$ can be calculated by
Landau's method. The expression of $F$ was given in Table I in
Ref. \cite{MSW7} and Table III in Ref. \cite{MSW8} for different
density distributions. Consequently, if we consider the following
electron density distribution
\begin{eqnarray}
 N_e=\frac{1}{m_n+m_p}\rho\simeq\frac{1}{2m_n} kr^n,
 \label{SNe}
\end{eqnarray}
where $k$ is a constant and $n$ is an integer, then
\begin{eqnarray}
\gamma=\frac{1}{2|n|}\Bigg(\frac{|\Delta
m^2|}{E}\Bigg)^{1+\frac{1}{n}}\Bigg(\frac{\sin^22\beta}{\cos2\beta}\Bigg)
\Bigg(\frac{\cos2\beta}{2\sqrt{2}G_F\frac{1}{2m_n}k}\Bigg)^{\frac{1}{n}},
\end{eqnarray}
\begin{eqnarray}
F=2\sum^{\infty}_{m=0}\left( \begin{array}{c}
 ~1/n-1 \\
 ~2m
\end{array} ~\right) \ \left[ \begin{array}{c}
 ~1/2 \\
 ~m+1
\end{array} ~\right] \ (\tan2\beta)^{2m},
\label{SF}
\end{eqnarray}
where
\begin{eqnarray}
 \left( \begin{array}{c}
 ~1/n-1 \\
 ~2m
\end{array} ~\right) \ &=& \frac{(1/n-1)!}{(1/n-1-2m)!(2m)!},
\nonumber
\end{eqnarray}
\begin{eqnarray}
 2\left[ \begin{array}{c}
 ~1/2 \\
 ~m+1
\end{array} ~\right] \ &=& (-1)^m\frac{J_m-J_{m+1}}{\pi/4},
\nonumber
\end{eqnarray}
\begin{eqnarray}
 J_m=\int^{\frac{\pi}{2}}_0(\sin\phi)^{2m}d\phi=\frac{(2m-1)!!}{(2m)!!}\frac{\pi}{2}.
 \label{correction}
\end{eqnarray}
%We note that our expression for $J_m$ in Eq. (\ref{correction}) is
%different from that given in Eq. (B5) in Appendix B of Ref.
%\cite{MSW7} which is not correct.
We note that our expression for $J_m$ in Eq. (\ref{correction})
has the same final result as given in Eq. (B5) in Appendix B of
Ref. \cite{MSW7}. But the defining integrals are different.

In order to obtain the expression for the crossing probability at
the high resonance region inside the SN, $P_H$, we can use the
similar method in the case of three-flavor transitions
\cite{PH1}\cite{PH2}. It is known that the following matter
density profile for the SN is appropriate
\cite{SNsum}\cite{SN1}\cite{SN2}\cite{SN3}:
\begin{equation}
\rho\approx C\cdot \Bigg(\frac{10^7cm}{r}\Bigg)^3\cdot10^{10}\quad
g/cm^3, \label{rho}
\end{equation}
where the constant $C$ depends on the amount of electron capture
during the star collapse and its value is between $1$ and $15$.
This corresponds to $n=-3$ and $k=C\cdot10^{31}$ in Eq.
(\ref{SNe}), then we have
\begin{eqnarray}
 \gamma= \frac{1}{6}\Bigg[\frac{10^{10} MeV}{E}\Bigg(\frac{\sin^32\theta_{13}}{\cos^22\theta_{13}}\Bigg)
 \Bigg(\frac{|\Delta m^2_{31}|}{1 eV^2}\Bigg)C^{1/2}\Bigg]^{2/3}.
 \nonumber
\end{eqnarray}
For very small $\theta_{13}$, $P_H$ has a simpler expression:
\begin{equation}
P_H=\exp(-\frac{\pi}{2}\gamma F). \nonumber
\end{equation}
Since $\theta_{13}$ is small, we need only to consider the $m=0$
term in Eq. (\ref{SF}), then
\begin{eqnarray}
 F\approx2\left( \begin{array}{c}
 ~1/n-1 \\
 ~0
\end{array} ~\right) \ \left[ \begin{array}{c}
 ~1/2 \\
 ~1
\end{array} ~\right] \ %(\tan2\theta_{13})^{0}
                           =1.
\nonumber
\end{eqnarray}
In addition, since $\Delta m^2_{21}\ll |\Delta m^2_{31}|$, we can
set %it can be seen that
\begin{eqnarray}
 |\Delta m^2_{31}|\thickapprox |\Delta m^2_{32}|.\nonumber
 \end{eqnarray}
Then we obtain
\begin{eqnarray}
P_H=\exp\Bigg\{-\frac{\pi}{12}\Bigg[\frac{10^{10}
MeV}{E}\Bigg(\frac{\sin^32\theta_{13}}{\cos^22\theta_{13}}\Bigg)
 \Bigg(\frac{|\Delta m^2_{32}|}{1 eV^2}\Bigg)C^{1/2}\Bigg]^{2/3}\Bigg\}. \label{phthree}
\end{eqnarray}
In the $2\sigma$ allowed ranges
\cite{Fogli2}\cite{Fogli3}\cite{Fogli4}\cite{SK1}
\begin{eqnarray}
 |\Delta m^2_{32}|&=&2.6\times(1^{+0.14}_{-0.15})\times10^{-3} eV^2 ,\nonumber\\
 \theta_{13}&<&10^\circ,
\end{eqnarray}
and we will use $2.6\times10^{-3}eV^2$ for $|\Delta m^2_{32}|$ in
the following calculation.

From Eq. (\ref{phthree}), the dependence of $P_H$ on $\theta_{13}$
and $E$ is shown explicitly. In order to obtain such simple
expression of $P_H$, we assume $\theta_{13}$ to be very small and
use the simplified structure model of SN in which $\rho\propto
r^{-3}$ in the above calculations. Therefore, Eq. (\ref{phthree})
is usually suitable for very small $\theta_{13}$ (we have checked
numerically that in the range of $\theta_{13}$ we are working Eq.
(\ref{phthree}) is a good approximation), and
 %%% someone needs to calculate
the expression for $P_H$ should be reevaluated if $\theta_{13}$ is
larger.

In Fig. 5(a), we plot $P_H$ as a function of $\theta_{13}$ and
 %%%
it can be seen that when $\theta_{13}$ is zero, $P_H=1$; in the
small range of $\theta_{13}$ of $0^{\circ}\thicksim2^{\circ}$
$P_H$ varies rapidly in the range $0<P_H<1$;
 %%%$$0^{\circ}\thicksim2^{\circ}$; , $\theta_{13}$ varies rapidly in
 %%%the very small range, $0^{\circ}\thicksim2^{\circ}$;
and when $\theta_{13}$ is larger than $2^\circ$, $P_H=0$.
Therefore, it is possible to use the value of $P_H$ to obtain the
value of $\theta_{13}$ when it is in the range $0^\circ\sim
2^\circ$. In Fig. 5(b), we plot $P_H$ as a function of $E$ with
$\theta_{13}$ in the range $0^\circ\sim 2^\circ$ and again it
shows that $\theta_{13}$ can be measured by the energy spectrum of
$P_H$ if the value of $\theta_{13}$ is in the range of
$0^\circ\sim 2^\circ$.

Using Eqs. (\ref{Ntotal}), (\ref{Foa2}), (\ref{FDN}), (\ref{FDI}),
and (\ref{phthree}), we obtain the relation between the event
number of SN neutrinos $N$ and the neutrino mixing angle
$\theta_{13}$. The curves of $N$ versus $\theta_{13}$ in the range
$0^{\circ}\thicksim3^{\circ}$ for the Daya Bay experiment for the
inverse beta-decay is given in Fig. 6. From this plot, we can see
that when $\theta_{13}\leq1.5^\circ$, $N$ is very sensitive to
$\theta_{13}$. When $\theta_{13}>1.5^\circ$, $N$ is nearly
independent of $\theta_{13}$. Therefore, when $\theta_{13}$ is
smaller than $1.5^{\circ}$, one could acquire some information
about $\theta_{13}$ by detecting the event numbers of SN
neutrinos. However, in practice, since we do not have the accurate
values of the SN neutrino flux parameters $T_{\alpha}$ and
$\eta_{\alpha}$ and they vary in some ranges, the event number of
SN neutrinos which we will detect depend on these two unknown
parameters strongly. It can be seen from Fig. 6 that the
uncertainties of the event number due to those of $T_{\alpha}$ and
$\eta_{\alpha}$ are very large, and it is difficult to obtain the
value of $\theta_{13}$ from the event number of SN neutrinos. This
difficulty also occurs in the neutrino-electron scattering and
neutrino-carbon reactions. Therefore, we need to work with a
quantity which changes with $\theta_{13}$ but is insensitive to
the values of $T_{\alpha}$ and $\eta_{\alpha}$.

In order to reduce uncertainties from unknown quantities such as
luminosity \cite{Janka2}, distance \cite{Ahrens}, temperatures
$T_{\alpha}$, and pinching parameters $\eta_{\alpha}$, ratios of
measurable quantities can be used.  For instance, \cite{raio}
defines the ratio of high-energy to low-energy event numbers that
are measurable in neutrino oscillation experiments:
\begin{eqnarray}
 R=\frac{N_{event}(30<E<70MeV)}{N_{event}(5<E<20MeV)}.\label{ratio}
\end{eqnarray}
Also, many different quantities to obtain the information about
$\theta_{13}$ and the mass hierarchy are suggested in \cite{PH1}.
In the following we choose a suitable reaction and then define a
ratio of event numbers which is sensitive to the mixing angle
$\theta_{13}$ but only dependents on $T_{\alpha}$ and
$\eta_{\alpha}$ slightly.

For the inverse beta-decay, there is only one flavor neutrino
$\bar{\nu}_e$. The numerator and denominator or $R$ of Eq.
(\ref{ratio}) have the same flavor and our calculation shows that
it has a significant dependence on $T_\alpha$ and $\eta_\alpha$.
Hence $R$ is not a suitable quantity to work with. In the case of
neutrino-electron scattering, the event number of SN neutrinos
detected at the Daya Bay experiment is very small as shown in the
above section. Again this channel of reaction is not useful for
our purpose.  We now turn our attention to the channel of the
neutrino-carbon reactions at the Daya Bay experiment.

For the neutrino-carbon reactions, we define the quantity $R$ to
be the ratio which is the event number of $\nu_{e}$ over that of
$\bar{\nu}_e$.  Using Eqs. (\ref{Ntotal}), (\ref{Foa2}),
(\ref{FDN}), (\ref{FDI}), and (\ref{phthree}), we can obtain the
relation between $R$ and the mixing angle $\theta_{13}$. With
this, we can obtain some information about $\theta_{13}$ smaller
than $1.5^\circ$ since the uncertainties due to $T_{\alpha}$ and
$\eta_{\alpha}$ are small. In Fig. 7, we plot $R$ as a function of
the mixing angle $\theta_{13}$ when $T_{\alpha}$ and
$\eta_{\alpha}$ take their limiting values in the ranges given in
(\ref{T}) and (\ref{eta}) for different incident angle $\theta$.
It can be seen from these plots that the uncertainties of $R$ due
to $T_{\alpha}$ and $\eta_{\alpha}$ are indeed not large. For
$\theta_{13}\leq1.5^\circ$, $R$ is very sensitive to
$\theta_{13}$. However, while $\theta_{13}>1.5^\circ$, $R$ is
nearly independent of $\theta_{13}$. Therefore, when $\theta_{13}$
is smaller than $1.5^{\circ}$, we may restrict the mixing angle
$\theta_{13}$ in a small range and get information about mass
hierarchy by detecting the ratio of event numbers of SN neutrinos
even though there are still some uncertainties due to the incident
angle $\theta$, the mass hierarchy $\triangle m_{31}^2$, and the
structure coefficient $C$ of the SN density function as given in
Eq. (\ref{rho}). Recently, there are discussions on methods for
the determination of the neutrino mass hierarchy and the incident
angle of the SN. For examples, in Ref. \cite{Dasgupta2}, a method
to identify the mass hierarchy at extremely small $\theta_{13}$
through the Earth matter effects is given, and a method to
determine the incident angle of the SN by the electron scattering
events can be found in \cite{Beacom1}. For the structure
coefficient $C$, its value can be more precisely determined if we
can obtain more exact density profile of the SN \cite{SN3}. In the
future, if the incident angle $\theta$, the mass hierarchy
$\triangle m_{31}^2$, and the the structure coefficient $C$ can be
determined, the uncertainties in the determination of
$\theta_{13}$ through the SN neutrino will be much reduced.

At the Daya Bay experiment,
 %%% the data taking is planned to begin in
 %%% 2009 and will last for three years. In the end,
the sensitivity of $\sin^22\theta_{13}$ will reach 0.01, i.e., to
determine $\theta_{13}$ down to about $3^\circ$.
%In other words, if
Therefore, if the actual value of $\theta_{13}$ is smaller than
$3^\circ$, the Daya Bay experiment can only provide an upper limit
for $\theta_{13}$. However, if an SN explosion takes place during
the operation of Daya Bay, roughly within the cosmic distance
considered here, it is possible to reach a much smaller value of
$\theta_{13}$ through the ratio of the event number of different
flavor SN neutrinos in the channel of neutrino-carbon reactions as
discussed above.

\section{\label{sec:cpv1}Acquiring information of $\theta_{13}$ smaller than $1.5^\circ$ from other neutrino experiments}

 %%% In the above section, we have proposed a method to determine very
 %%% small $\theta_{13}$ by
%detecting the event number of
 %%% the SN neutrinos and applied it to the Daya Bay experiment.
In this section, we will apply the above method to some other
current neutrino experiments including Super-K, SNO, KamLAND, LVD,
MinBooNE, Borexino, and Double-Chooz which is under construction.

For Super-K, the target material is water, the total detector mass
is 32000 tons, and the depth of the detector $h=2700$
m.w.e.\footnote{m.w.e. refers to meter-water-equivalent and 1 m of
rock is about 2.7 m of water.}
\cite{SK1}\cite{raio}\cite{Beacom2}\cite{SK2}. Then the total
numbers of the targets (protons, electrons, and $^{16}O$) are
\begin{eqnarray}
N_T^{(p)}=2.14\times10^{33},  \nonumber\\
N_T^{(e)}=1.07\times10^{34},  \nonumber\\
N_T^{(O)}=1.07\times10^{33}. \label{NC}
\end{eqnarray}
First, we consider the inverse beta-decay. When the incident angle
$\theta=30^\circ$, we use Eqs. (\ref{Ntotal}), (\ref{Foa2}),
(\ref{FDN}), (\ref{FDI}), and (\ref{phthree}) to plot the event
number of SN neutrinos at Super-K as a function of the mixing
angle $\theta_{13}$. The result is shown in Fig. 8. Similar to the
Daya Bay experiment, it can be seen that the uncertainties of the
event number due to the two quantities $T_{\alpha}$ and
$\eta_{\alpha}$ are very large, and hence we cannot make
predictions on $\theta_{13}$ from the event number of SN
neutrinos. Therefore, we need to choose a suitable reaction at
Super-K and find out a quantity which is sensitive to
$\theta_{13}$ but only dependents on $T_{\alpha}$ and
$\eta_{\alpha}$ slightly.

Although the event number of SN neutrinos in the channel of
neutrino-electron scattering is very small at Daya Bay, it is much
larger at Super-K due to the large number of target electrons
$N_T^{(e)}$.  We define $R$ as the ratio of the event number of
$\nu_{e}$ to that of $\bar{\nu}_e$ from SN. Using Eqs.
(\ref{Ntotal}), (\ref{Foa2}), (\ref{FDN}), (\ref{FDI}), and
(\ref{phthree}), we can obtain the relation between $R$ and the
mixing angle $\theta_{13}$. By measuring this ratio, we can
acquire some useful information about $\theta_{13}$ smaller than
$1.5^\circ$ since the uncertainties due to $T_{\alpha}$ and
$\eta_{\alpha}$ are small. In Fig. 9, we plot $R$ as a function of
the mixing angle $\theta_{13}$ when $T_{\alpha}$ and
$\eta_{\alpha}$ take their limiting values in the ranges (\ref{T})
and (\ref{eta}) for different incident angle $\theta$. It can be
seen from these plots that the uncertainties in $R$ due to
$T_{\alpha}$ and $\eta_{\alpha}$ are not large. When
$\theta_{13}\leq1.5^\circ$, $R$ is very sensitive to
$\theta_{13}$. However, while $\theta_{13}>1.5^\circ$, $R$ is
nearly independent of $\theta_{13}$. Therefore, when $\theta_{13}$
is smaller than $1.5^{\circ}$, $\theta_{13}$ can be constrained in
a small range by the event numbers of SN neutrinos. Furthermore,
in spite of uncertainties due to $\theta_{13}$, $T_{\alpha}$,
$\eta_{\alpha}$ and $C$, the plots show that if $R$ is smaller
than about $2.35$, the mass hierarchy must be normal while if $R$
is larger than about $2.5$, the mass hierarchy must be inverted.
%Therefore, we may also determine
%the mass hierarchy by the ratio of the event numbers of $\nu_{e}$
%and $\bar{\nu}_e$ which we can detect from SN in the channel of
%neutrino-electron scattering.

For SNO,  the detector material was heavy water, the total
detector mass was 1000 tons, and the depth of the detector
$h=6000$ m.w.e. \cite{SNO1}\cite{Beacom3}. Therefore, we need to
consider the reactions between neutrinos and deuterium. There are
two changed-current and six neutral-current reactions. The cross
sections are obtained by scaling the experimentally measured
energy values from the decay of the muon at rest to the energy
scale for SN neutrinos. In this way, one can obtain
\cite{Burrows1}\cite{raio}\cite{Burrows3}\cite{Cross-Section}:

Charged-current capture of $\nu_e$ or $\Bar{\nu}_e$:
\begin{eqnarray}
&&\ \nu_e+d\rightarrow p+p+e^-, \quad E_{th}=1.44MeV, \nonumber
\\
 &&\ \Bar{\nu}_e+d\rightarrow n+n+e^+, \quad E_{th}=4.03MeV,
\nonumber \\
&&\
\langle\sigma(d(\nu_e,e^{-})pp)\rangle=(3.35T_{\nu_e}^{2.31}-3.70)\times10^{-43}
cm^2, \nonumber \\
&&\
 \langle\sigma(d(\bar{\nu}_e,e^{+})nn)\rangle=(3.05T_{\bar{\nu}_e}^{2.08}-7.82)\times10^{-43}
cm^2. \nonumber
\end{eqnarray}

Neutral-current inelastic scattering of $\nu_{\alpha}$ or
$\Bar{\nu}_{\alpha}$ where $\alpha=e, \mu, \tau$:
\begin{eqnarray}
&&\ \nu_{\alpha}+d\rightarrow n+p+\nu_{\alpha}^{'}, \quad
E_{th}=2.22MeV,
\nonumber \\
&&\ \bar{\nu}_{\alpha}+d\rightarrow n+p+\bar{\nu}_{\alpha}^{'},
\quad E_{th}=2.22MeV,
\nonumber \\
 &&\ \langle\sigma(d(\nu_{\alpha},\nu_{\alpha}^{'})np)\rangle=(1.63T_{\nu_{\alpha}}^{2.26}-2.78)\times10^{-43}
cm^2, \nonumber \\
 &&\ \langle\sigma(d(\bar{\nu}_{\alpha},\bar{\nu}_{\alpha}^{'})np)\rangle=
 (2.03T_{\bar{\nu}_{\alpha}}^{2.05}-3.76)\times10^{-43}cm^2. \nonumber
\end{eqnarray}

For $T_{\nu_e}=3.5MeV$, $T_{\bar{\nu}_e}=5MeV$, $T_{\nu_x}=8MeV$,
\begin{eqnarray}
&&\ \langle\sigma(d(\nu_e,e^{-})pp)\rangle=5.68\times10^{-42}
cm^2, \nonumber\\
&&\
 \langle\sigma(d(\bar{\nu}_e,e^{+})nn)\rangle=7.89\times10^{-42}
cm^2, \nonumber\\
&&\
\langle\sigma(d(\nu_e,\nu_{\alpha}^{'})np)\rangle=2.49\times10^{-42}
cm^2, \nonumber \\
 &&\ \langle\sigma(d(\bar{\nu}_e,\bar{\nu}_{\alpha}^{'})np)\rangle=
 5.12\times10^{-42}cm^2, \nonumber \\
&&\ \langle\sigma(d(\nu_x,\nu_x^{'})np)\rangle=1.76\times10^{-41}cm^2, \nonumber \\
 &&\
 \langle\sigma(d(\bar{\nu}_x,\bar{\nu}_x^{'})np)\rangle=1.40\times10^{-41}cm^2, \quad x=\mu,\tau.
 \label{sigma D}
\end{eqnarray}
As indicated in Eq. (\ref{sigma D}), it can be seen that the
average effective cross sections of $\nu_x$ and $\bar{\nu}_x$
reactions are different. Therefore, we need to distinguish the
energy spectrum functions $F_{\nu_x}^{(0)}$ and
$F_{\bar{\nu}_x}^{(0)}$ appearing in Eqs. (\ref{FDN}) and
(\ref{FDI}).

Similar to the neutrino-carbon scattering, the average effective
cross sections in Eq. (\ref{sigma D}) are given for SN neutrinos
without oscillations \cite{Cross-Section}. When neutrino
oscillations are taken into account, the oscillation of higher
energy $\nu_x$ into $\nu_e$ again results in an increased event
rate since the expected $\nu_e$ energies are just at or below the
charged-current reaction threshold. This leads to an increase by a
factor of 7 for the cross section
$\langle\sigma(d(\nu_e,e^{-})pp)\rangle$. Similarly, the cross
section $\langle\sigma(d(\bar{\nu}_e,e^{+})nn)\rangle$ is
increased by a factor of 3. For the case of neutral-current
inelastic scattering of $\nu_{\alpha}$ or $\Bar{\nu}_{\alpha}$,
when the oscillation of higher energy $\nu_x$ into $\nu_e$ is
taken into account, the cross section
$\langle\sigma(d(\nu_e,\nu_{\alpha}^{'})np)\rangle$ is increased
by a factor of 7, while when the oscillation of higher energy
$\nu_x$ into $\Bar{\nu}_e$ is considered, the cross section
$\langle\sigma(d(\bar{\nu}_e,\bar{\nu}_{\alpha}^{'})np)\rangle$ is
increased by a factor of 3.

Now, we define $R$ as the ratio of the event number of $\nu_{e}$
to that of $\bar{\nu}_e$ in the channel of neutrinos-deuterium
scattering. Using Eqs. (\ref{Ntotal}), (\ref{Foa2}), (\ref{FDN}),
(\ref{FDI}), and (\ref{phthree}), we can obtain the relation
between $R$ and the mixing angle $\theta_{13}$. In Fig. 10, we
plot $R$ as a function of $\theta_{13}$ when $T_{\alpha}$ and
$\eta_{\alpha}$ take their limiting values in (\ref{T}) and
(\ref{eta}) for different incident angle $\theta$. Similar to
electron-carbon reactions, it can be seen from these plots that
the uncertainties of $R$ due to $T_{\alpha}$ and $\eta_{\alpha}$
are not large. When $\theta_{13}\leq1.5^\circ$, $R$ is very
sensitive to $\theta_{13}$. However, while
$\theta_{13}>1.5^\circ$, $R$ is nearly independent of
$\theta_{13}$. Therefore, when $\theta_{13}$ is smaller than
$1.5^{\circ}$, we can constrain $\theta_{13}$ in a small range by
the detection of the event numbers of SN neutrinos in the channel
of neutrinos-deuterium reactions at SNO.

Similar to the Daya Bay experiment, the method to obtain
information about $\theta_{13}$ smaller than $1.5^\circ$ can be
applied to KamLAND \cite{KamLAND}, LVD \cite{LVD}, MinBooNE
\cite{MinBooNE}, Borexino \cite{Borexino}, and Double-Chooz
\cite{Chooz} in the channel of neutrino-carbon reactions. We can
also constrain $\theta_{13}$ in a small range with the ratio of
the event number of SN $\nu_{e}$ to that of $\bar{\nu}_e$ in these
experiments. Since the number of target carbons of Double-Chooz is
much smaller than that of Daya Bay, the event number of SN
neutrinos which can be detected is fewer and it will be more
difficult to obtain information about $\theta_{13}$ by our method.

\section{\label{sec:cpv1}Energy spectra of the differential event numbers}

In this section, we consider the energy spectra of the
differential event numbers, referred generically as ${dN}/{dE}$,
where $E$ is the neutrino energy. From Eq. (\ref{Ntotal}), one can
obtain
\begin{equation}
% \frac{{\rm d}N}{{\rm d}E}=N_T\cdot\sigma(i)\cdot\frac{1}{4\pi
% D^2}\cdot F_{\alpha}^D. \label{dNdE}
 \frac{{\rm d}N_\alpha(i)}{{\rm d}E}=N_T \sigma(i) \frac{1}{4\pi D^2}
        F_{\alpha}^D. \label{dNdE}
\end{equation}
Since the inverse beta-decay is the most important reaction among
the three kinds of reactions in the Daya Bay experiment, we
examine the energy spectra of this process, taking the simplest
parameter set $T_{\nu_e}=3.5MeV$, $T_{\bar{\nu}_e}=5MeV$,
$T_{\nu_x}=8MeV$, $\eta_{\nu_e}=\eta_{\bar{\nu}_e}=\eta_{\nu_x}=0$
as an example.

Using Eqs. (\ref{sigma p}), (\ref{Foa2}), (\ref{FDN}),
(\ref{FDI}), and (\ref{dNdE}), we make a three-dimensional plot of
${dN}/{dE}$ versus $E$ in the Daya Bay experiment for different
Earth incident angle $\theta$ in Fig. 11(a)
% We give the result in the cases where
for $\theta_{13}=0$ (inverted hierarchy). We can see that when the
incident $\theta$ increases the Earth matter effects become more
and more obvious. The curve of ${dN}/{dE}$ changing with $E$ is
very smooth when $\theta$ is small. However, it becomes
oscillatory when $\theta$ is greater than $90^{\circ}$. Therefore,
this shows that the Earth matter effects on ${dN}/{dE}$ need to be
considered in this case.

Using Eqs. (\ref{sigma p}), (\ref{Foa2}), (\ref{FDN}),
(\ref{FDI}), (\ref{phthree}), and (\ref{dNdE}), we plot
${dN}/{dE}$ as a function of $E$ for different $\theta_{13}$ in
Fig. 11(b) for the case of inverted hierarchy, where the structure
coefficient of SN $C=3$, and the Earth incident angle
$\theta=30^\circ$ are used. We can see that the energy spectrum
changes with $\theta_{13}$ in the range $0^\circ\sim1.5^\circ$.

\section{Summary and discussions}

In this paper, we have calculated the realistic Earth matter
effects in the detection of type II SN neutrinos at the Daya Bay
experiment under construction. It is found that the Earth matter
effects depend on the neutrino incident angle $\theta$, the
neutrino mass hierarchy $\Delta m_{32}^2$, the crossing
probability at the high resonance region inside the SN, $P_{H}$,
the neutrino temperature $T_\alpha$, the pinching parameter in the
neutrino energy spectra $\eta_\alpha$, and the collective effects
of neutrino-neutrino interactions in the SN. We have given the
event numbers that can be detected through the inverse beta-decay,
the neutrino-electron scattering, and the neutrino-carbon
scattering. We have studied the effects due to the variations of
the neutrino temperature $T_{\alpha}$ and the pinching parameter
$\eta_{\alpha}$ in the neutrino energy spectra.
% By measuring the
% event numbers in various channels one can obtain some information
% on the neutrino mass hierarchy $\Delta m_{32}^2$, the neutrino
% mixing angle $\theta_{13}$, the neutrino temperature $T_{\alpha}$,
% and the pinching parameter in the spectra $\eta_{\alpha}$.

Since neutrino flavor conversions inside the SN depend on the
neutrino mixing angle $\theta_{13}$, it is possible to get
information about $\theta_{13}$ by detecting SN neutrinos. In
fact, there have been some general discussions on this possibility
\cite{PH1}\cite{Dighe}. We have made concrete calculations and
given explicit event numbers by using the relation between the
event numbers of SN neutrinos $N$ and the neutrino mixing angle
$\theta_{13}$ under different scenarios of neutrino parameters.
For $\theta_{13}$ smaller than $1.5^{\circ}$, we propose an
approach to constrain the value of angle $\theta_{13}$ in a small
range and get information about mass hierarchy by measuring the
ratio of the event numbers of different flavors of SN neutrinos.
For the Daya Bay experiment, we choose the ratio of the event
number of $\nu_e$ to that of $\bar{\nu}_e$ in the channel of
neutrino-carbon reactions. We have also applied this method to
other neutrino detectors including Super-K, SNO, KamLAND, LVD,
MinBooNE, Borexino, and Double-Chooz. For the Super-K, the
suitable reaction is the neutrino-electron scattering due to its
large number of target electrons. We have also shown that the
neutrino-deuterium reactions at SNO may be used to acquire useful
information about $\theta_{13}$.

We have studied the energy spectra of the differential event
numbers, ${dN}/{dE}$. From the dependence of ${dN}/{dE}$ on the
neutrino energy $E$ for different Earth incident angle $\theta$,
we have found that when the Earth incident angle $\theta$
increases the Earth matter effects become more and more
pronounced.

There are still some uncertainties in our work. In our
calculations, we took the distance from the SN to the Earth to be
10 kpc, where the maximum of the progenitor population appears in
the Milky Way \cite{Bahcall}. The distribution of the SN
progenitors as a function of the distance to the Earth can be
found in Ref. \cite{Ahrens}.

We let the parameters in the neutrino energy spectra (the
temperatures and the pinching parameters) vary in some reasonable
ranges. In fact, the simulations from the two leading groups, the
Livermore group \cite{Totani} and the Garching group \cite{Janka2}
(which considered more reactions in their simulation and found
different dominant neutrino production processes in the formation
of the SN neutrino spectra and fluxes), led to parameters which
agree within about 20-30\%. However, their central values of SN
parameters are different.

\section{Acknowledgments}

We would like to thank J.-Sh. Deng, H.-L. Xiao, M.-J. Chen, K.-F.
Chen, M.-H. Weng, and X.-H. Wu for helpful discussions. This work
was supported in part by National Natural Science Foundation of
China (Project Numbers 10535050 and 10675022), the Key Project of
Chinese Ministry of Education (Project Number 106024) and the
Special Grants from Beijing Normal University. BLY would like to
thank Yue-Liang Wu and Jin Min Yang of the Institute of
Theoretical Physics and Xin-Heng Guo of Beijing Normal University
for their warm hospitality and support.

%\onecolumn

\newpage

\noindent{\large \bf Figure Captions} \\
\vspace{0.4cm}

\noindent Fig. 1 Illustration of the path of the SN neutrino
reaching the detector in the Earth. $D$ is the location of the
detector, $\theta$ is the incident angle of the neutrino,
 $O$ is the center of the Earth, $L$ is the distance the neutrino
travels through the Earth, and $\tilde{x}$ is the distance of the
neutrino to the center of the Earth.\vspace{0.2cm}

\noindent Fig. 2 The event numbers observed at the Daya Bay
experiment as a function of the incident angle $\theta$ when
$T_{\nu_e}=3.5MeV$, $T_{\bar{\nu}_e}=5MeV$, $T_{\nu_x}=8MeV$, and
$\eta_{\nu_e}=\eta_{\bar{\nu}_e}=\eta_{\nu_x}=0$. (a) the channel
$\bar{\nu}_e+p\rightarrow e^++n$; (b) the reactions
$\nu+e^-\rightarrow \nu+e^-$; (c) the neutrino-carbon reactions.
The solid curves correspond to $P_H=1$ (normal hierarchy), the
dashed curves correspond to $P_H=1$ (inverted hierarchy), the
dotted curves correspond to $P_H=0$ (normal hierarchy), and the
dot-dashed curves correspond to $P_H=0$ (inverted hierarchy). For
(a), the solid curve also corresponds to $P_H=1$ (inverted
hierarchies) and $P_H=0$ (normal hierarchy).\vspace{0.2cm}

\noindent Fig. 3 Similar to Fig. 2 but $T_{\alpha}$ and
$\eta_{\alpha}$ take their maximum values: $T_{\nu_e}=4MeV$,
$T_{\bar{\nu}_e}=6MeV$, $T_{\nu_x}=9MeV$, $\eta_{\nu_e}=5$,
$\eta_{\bar{\nu}_e}=2.5$, $\eta_{\nu_x}=2$. \vspace{0.2cm}

\noindent Fig. 4 Similar to Fig. 2 but $T_{\alpha}$ and
$\eta_{\alpha}$ take their minimum values: $T_{\nu_e}=3MeV$,
$T_{\bar{\nu}_e}=5MeV$, $T_{\nu_x}=7MeV$, $\eta_{\nu_e}=3$,
$\eta_{\bar{\nu}_e}=2$, $\eta_{\nu_x}=0$. \vspace{0.2cm}

\noindent Fig. 5 The crossing probability at the high resonance
region inside the SN: (a) as a function of the neutrino mixing
angle $\theta_{13}$ and the solid curves, dotted curves, dashed
curves correspond to $E$ = $11MeV$, $16MeV$, $25MeV$,
respectively; (b) as a function of the neutrinos energy $E$ and
the solid curves, dotted curves, dashed curves correspond to
$\theta_{13}$ = $0.1^\circ$, $1^\circ$, $2^\circ$, respectively.
\vspace{0.2cm}

\noindent Fig. 6 The event number observed in the channel
$\bar{\nu}_e+p\rightarrow e^++n$ at the Daya Bay experiment as a
function of the neutrino mixing angle $\theta_{13}$ when the
incident angle $\theta=30^\circ$. The solid curves correspond to
the normal hierarchy, and the dashed curves correspond to the
inverted hierarchy, where "max" ("min") corresponds to the maximum
(minimum) values of $T_{\alpha}$ and
$\eta_{\alpha}$.\vspace{0.2cm}

\noindent Fig. 7 The ratio of the event number of $\nu_e$ to that
of $\bar{\nu}_e$, $R$, as a function of the mixing angle
$\theta_{13}$ in the channel of neutrino-carbon reactions at the
Daya Bay experiment. (a) the incident angle $\theta=30^\circ$; (b)
$\theta=90^\circ$; (c) $\theta=93^\circ$; (d) $\theta=150^\circ$.
The solid curves correspond to the normal hierarchy (max), the
dashed curves correspond to the inverted hierarchy (max), the
dot-dashed curves correspond to the normal hierarchy (min), the
dotted curves correspond to the inverted hierarchy (min), where
"max" ("min") corresponds to the maximum (minimum) values of
$T_{\alpha}$ and $\eta_{\alpha}$.\vspace{0.2cm}

\noindent Fig. 8 Similar to Fig. 6 but for the Super-K
experiment.\vspace{0.2cm}

\noindent Fig. 9 Similar to Fig. 7 but for the ratio of the event
number of $\nu_e$ to that of $\bar{\nu}_e$, $R$, as a function of
the mixing angle $\theta_{13}$ in the channel of neutrino-electron
scattering at the Super-K experiment. \vspace{0.2cm}

\noindent Fig. 10 Similar to Fig. 7 for the ratio of the event
number of $\nu_e$ to that of $\bar{\nu}_e$, $R$, as a function of
the mixing angle $\theta_{13}$ in the channel of
neutrino-deuterium reactions at the SNO experiment. \vspace{0.2cm}

\noindent Fig. 11  In the case of inverted hierarchy, the
differential event number ${dN}/{dE}$ observed in the inverse
beta-decay channel at the Daya Bay experiment as a function of the
neutrino energy $E$. (a) for different incident angle $\theta$
($\theta_{13}=0$); (b) for different $\theta_{13}$
($\theta=30^\circ$, $C=3$) when $T_{\nu_e}=3.5MeV$,
$T_{\bar{\nu}_e}=5MeV$, $T_{\nu_x}=8MeV$,
$\eta_{\nu_e}=\eta_{\bar{\nu}_e}=\eta_{\nu_x}=0$.\vspace{0.2cm}

\newpage

\begin{figure}
\includegraphics[width=0.4\textwidth]{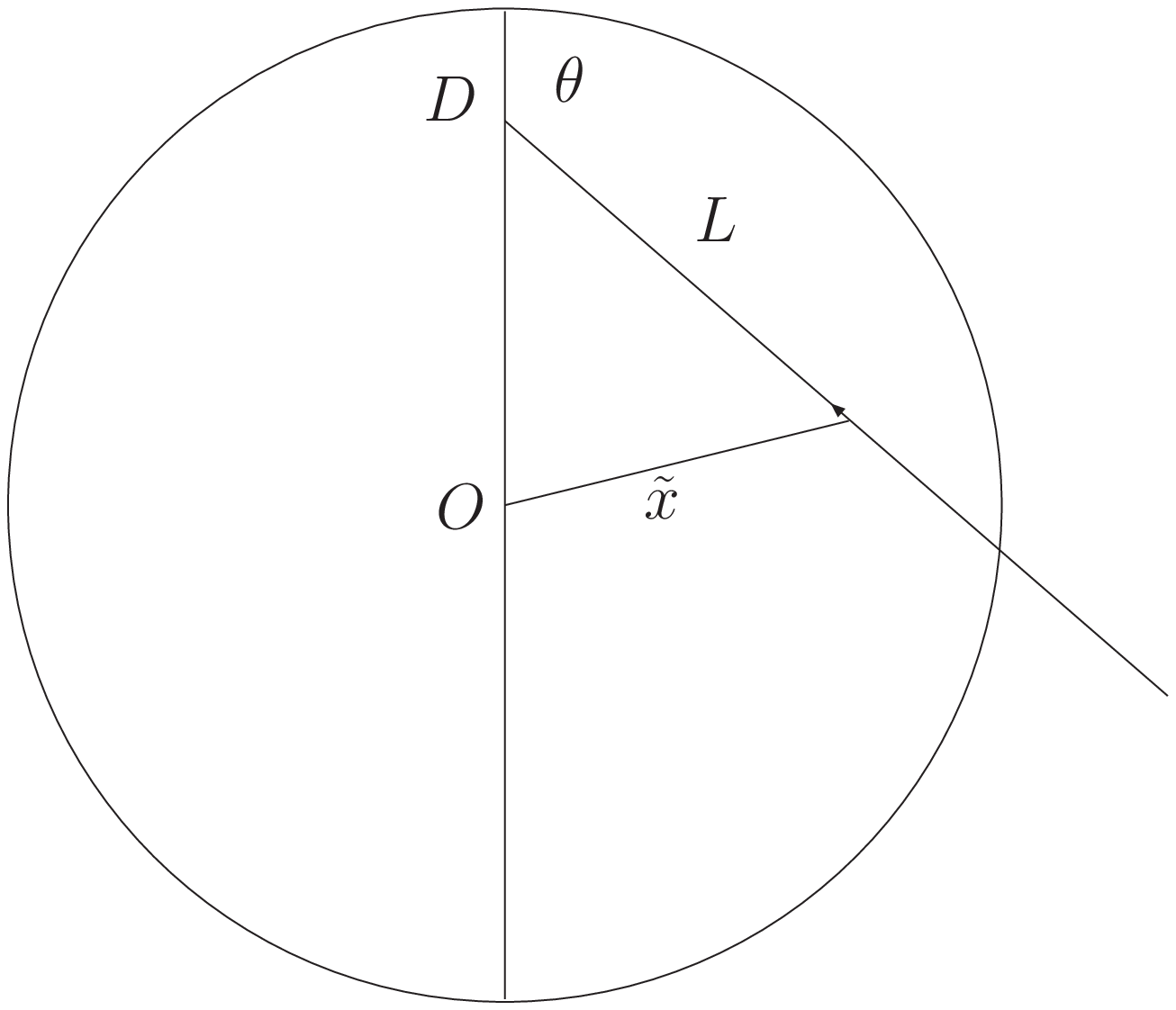}\\
\centerline{Fig. 1}
\end{figure}

\begin{figure}
\includegraphics[width=1\textwidth]{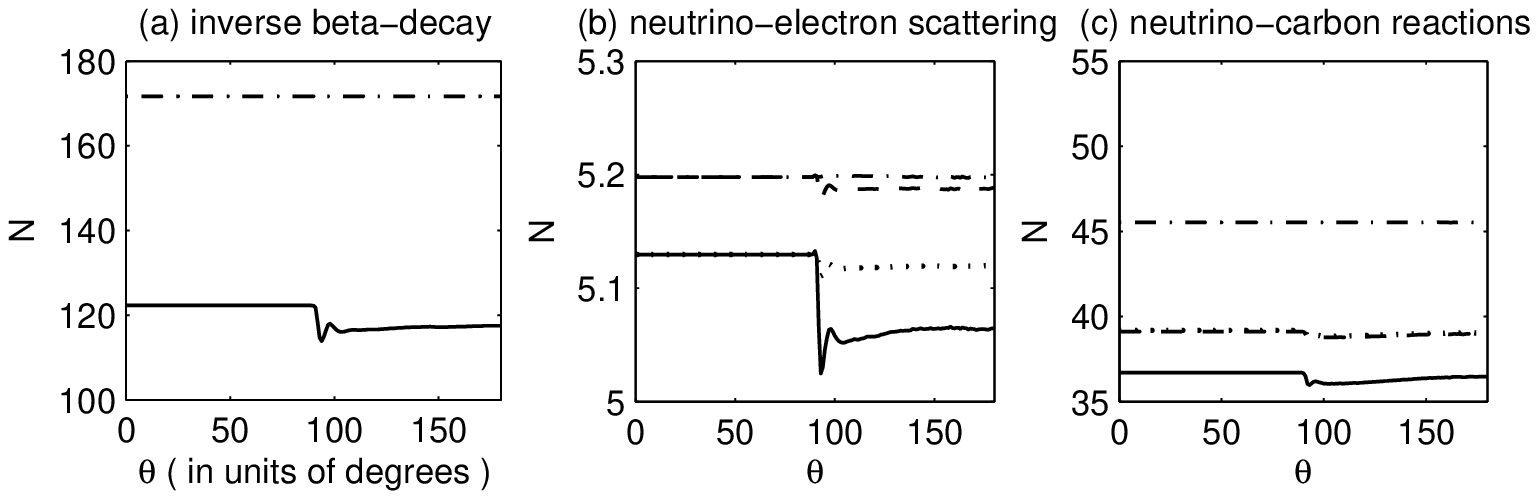}\\
\centerline{Fig. 2}
\end{figure}

\begin{figure}
\includegraphics[width=1\textwidth]{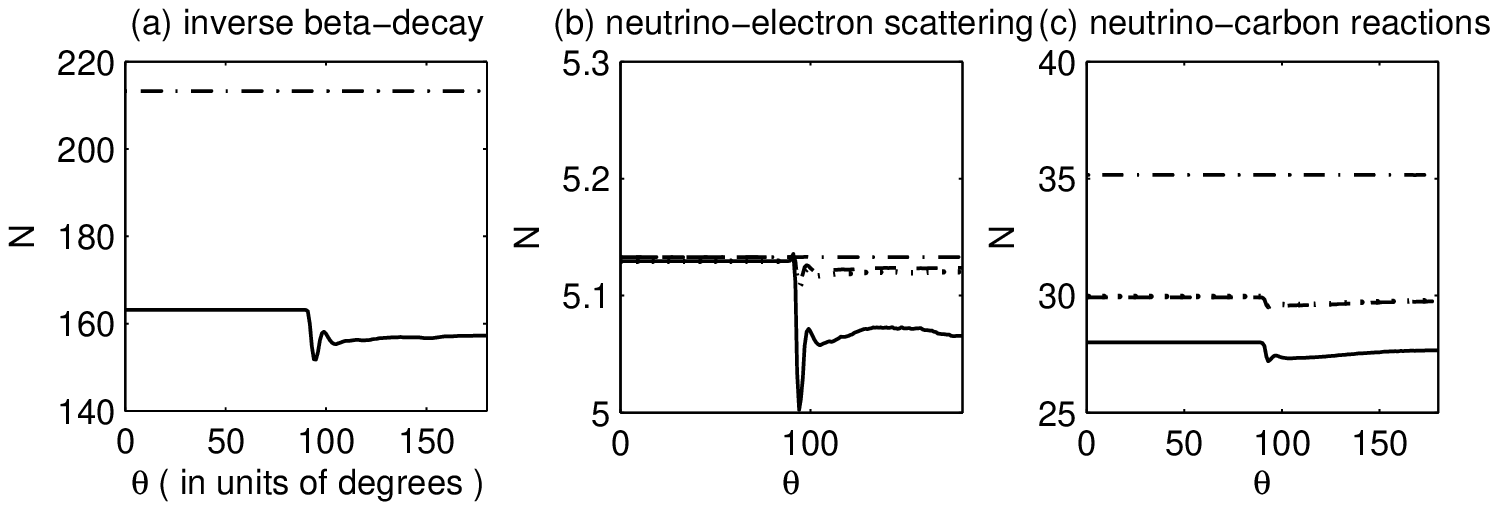}\\
\centerline{Fig. 3}
\end{figure}

\begin{figure}
\includegraphics[width=1\textwidth]{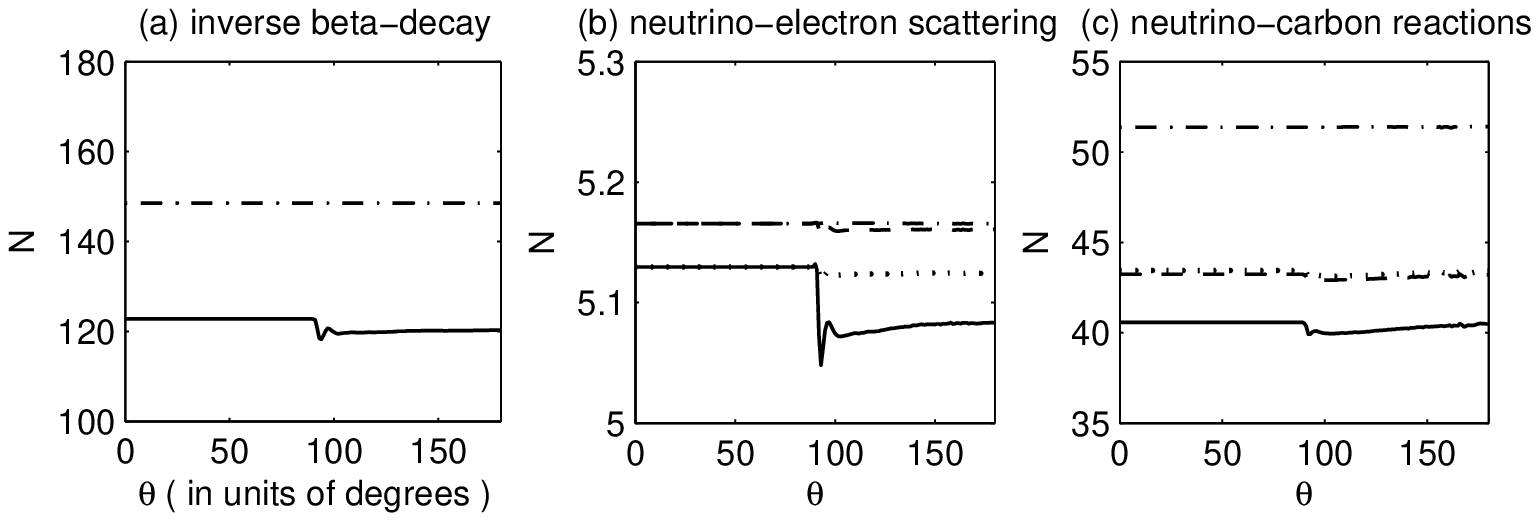}\\
\centerline{Fig. 4}
\end{figure}

\begin{figure}
\includegraphics[width=0.6\textwidth]{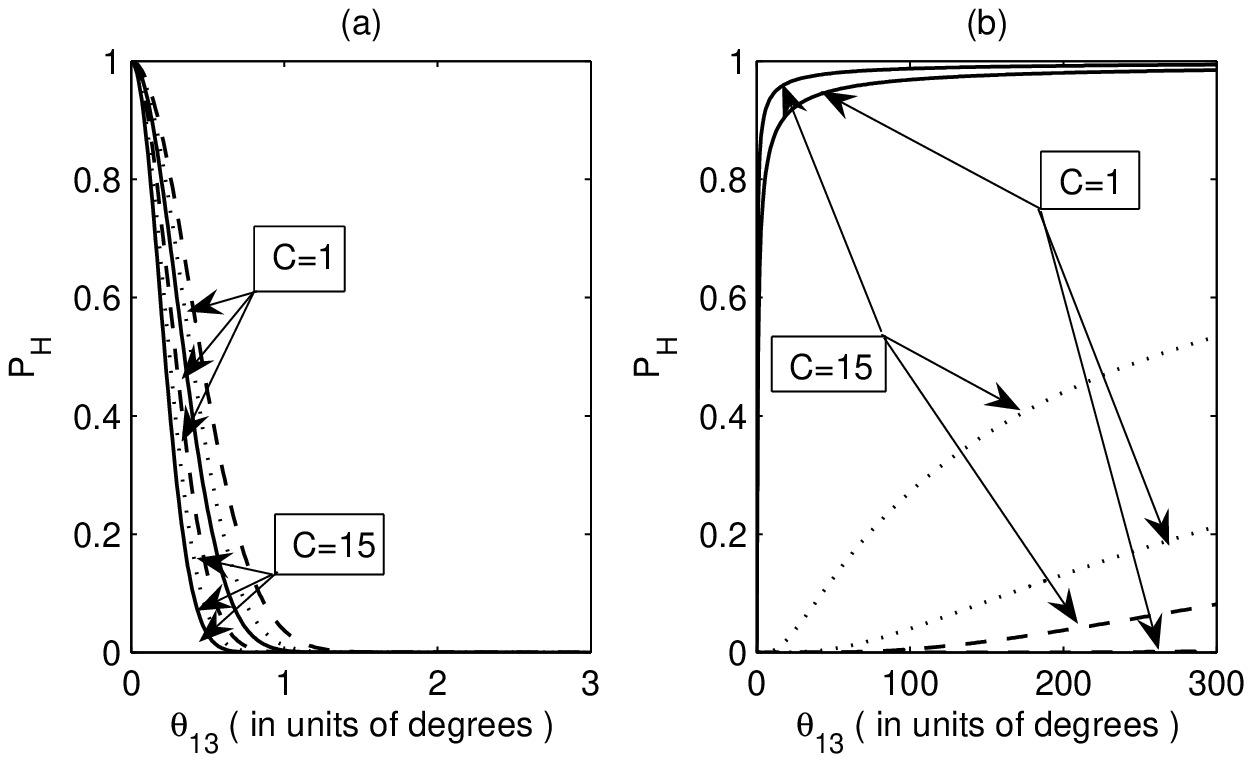}\\
\centerline{Fig. 5}
\end{figure}

\begin{figure}
\includegraphics[width=0.5\textwidth]{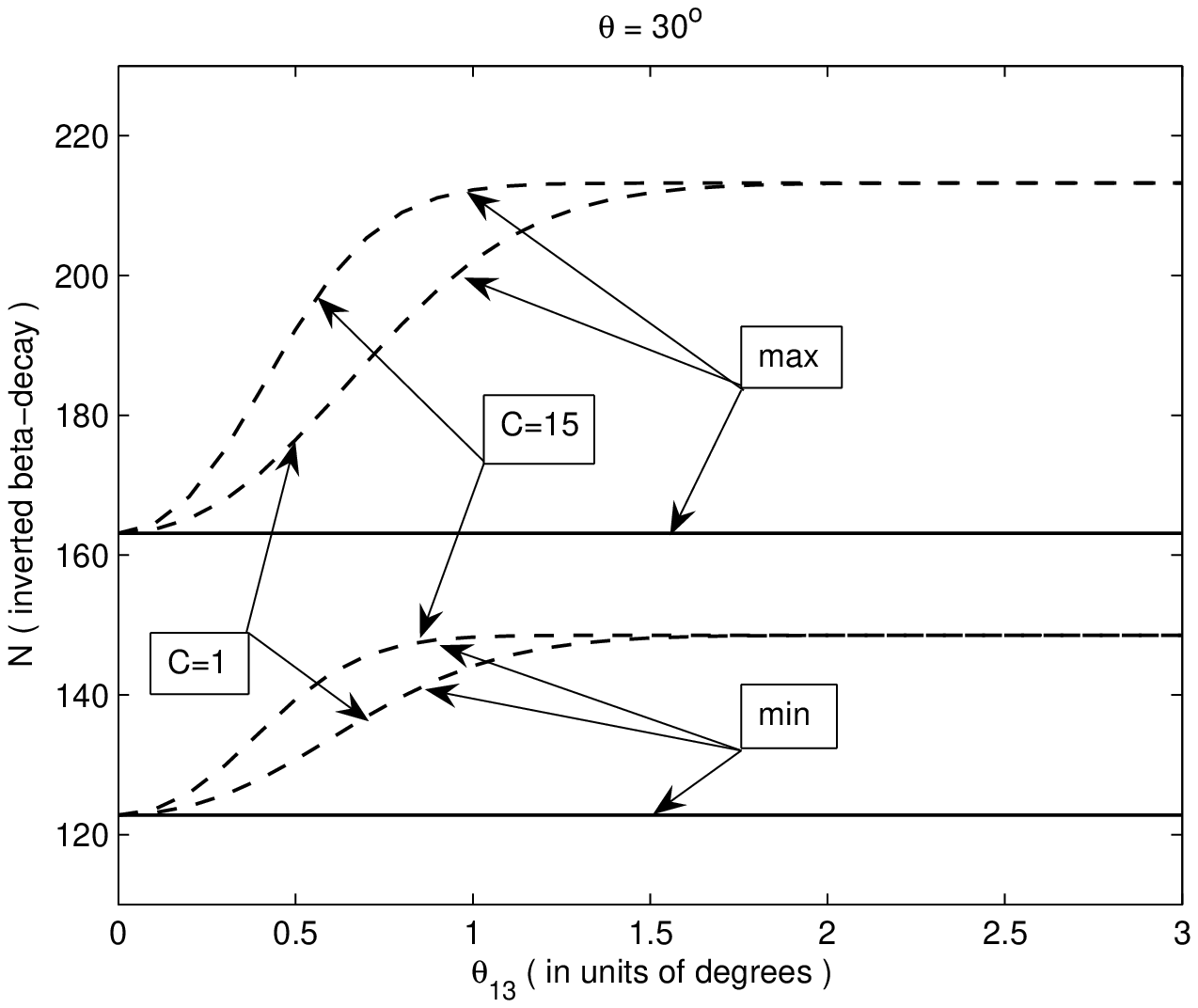}\\
\centerline{Fig. 6}
\end{figure}

\begin{figure}
\includegraphics[width=0.8\textwidth]{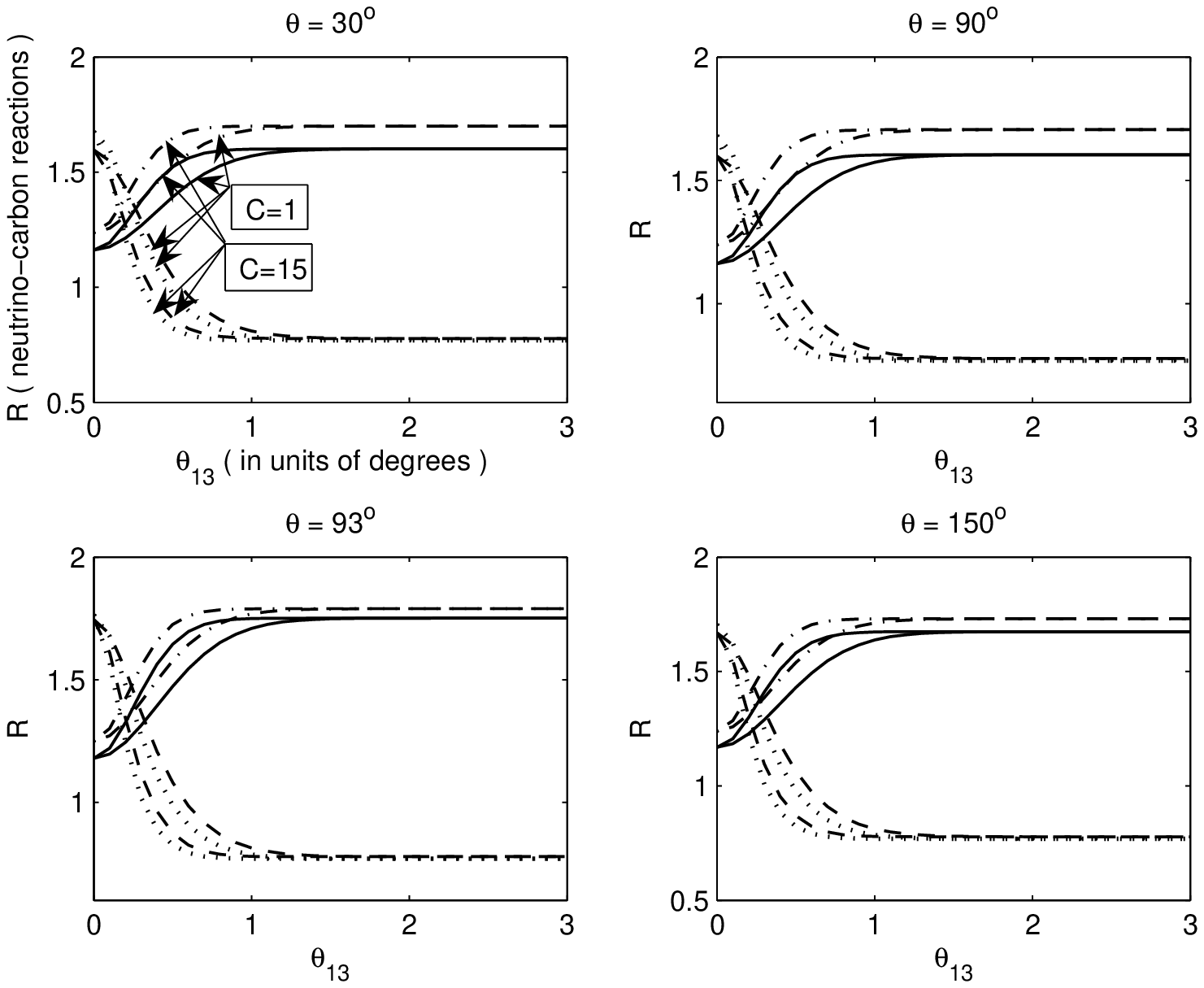}\\
\centerline{Fig. 7}
\end{figure}

\begin{figure}
\includegraphics[width=0.5\textwidth]{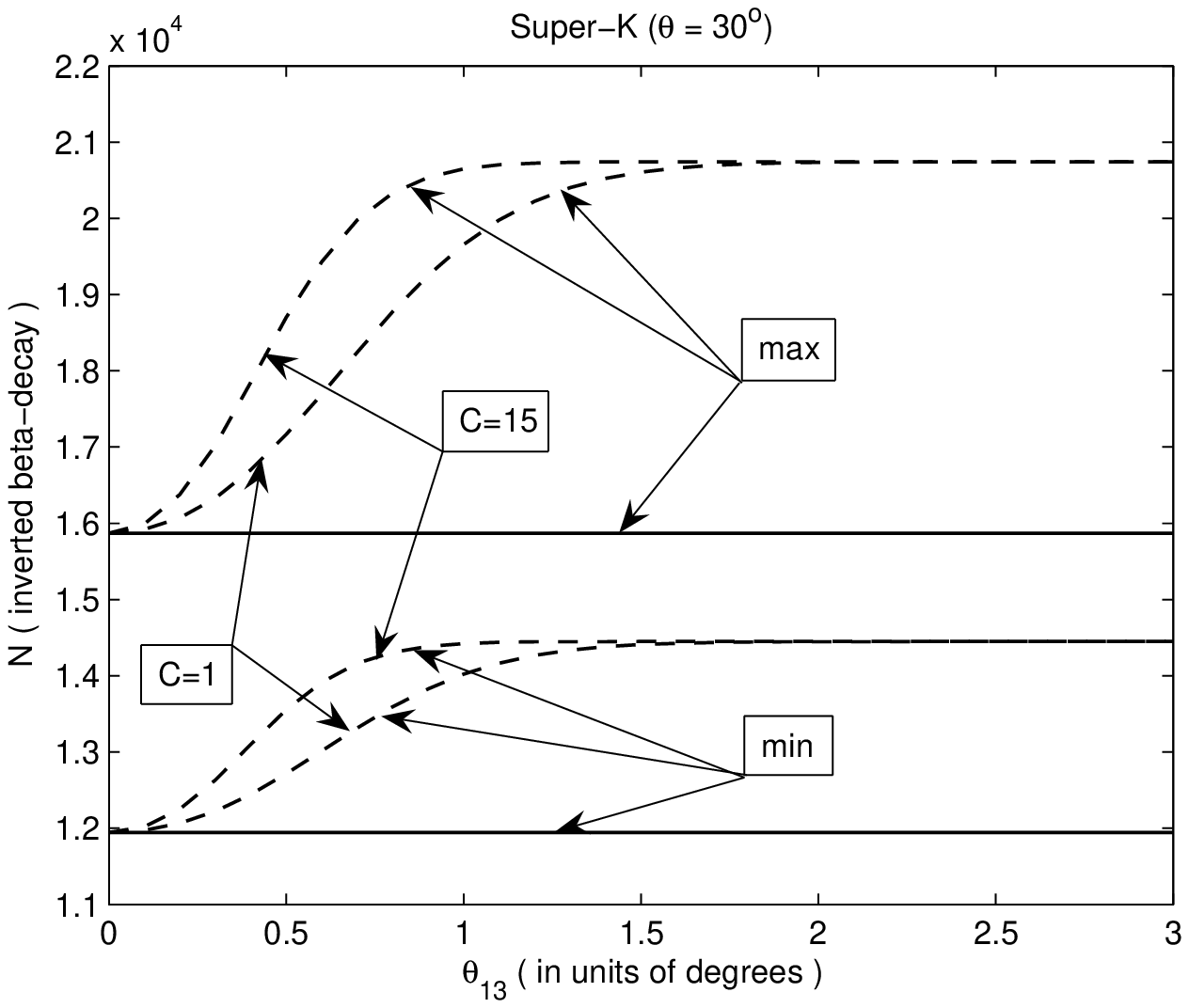}\\
\centerline{Fig. 8}
\end{figure}

\begin{figure}
\includegraphics[width=0.8\textwidth]{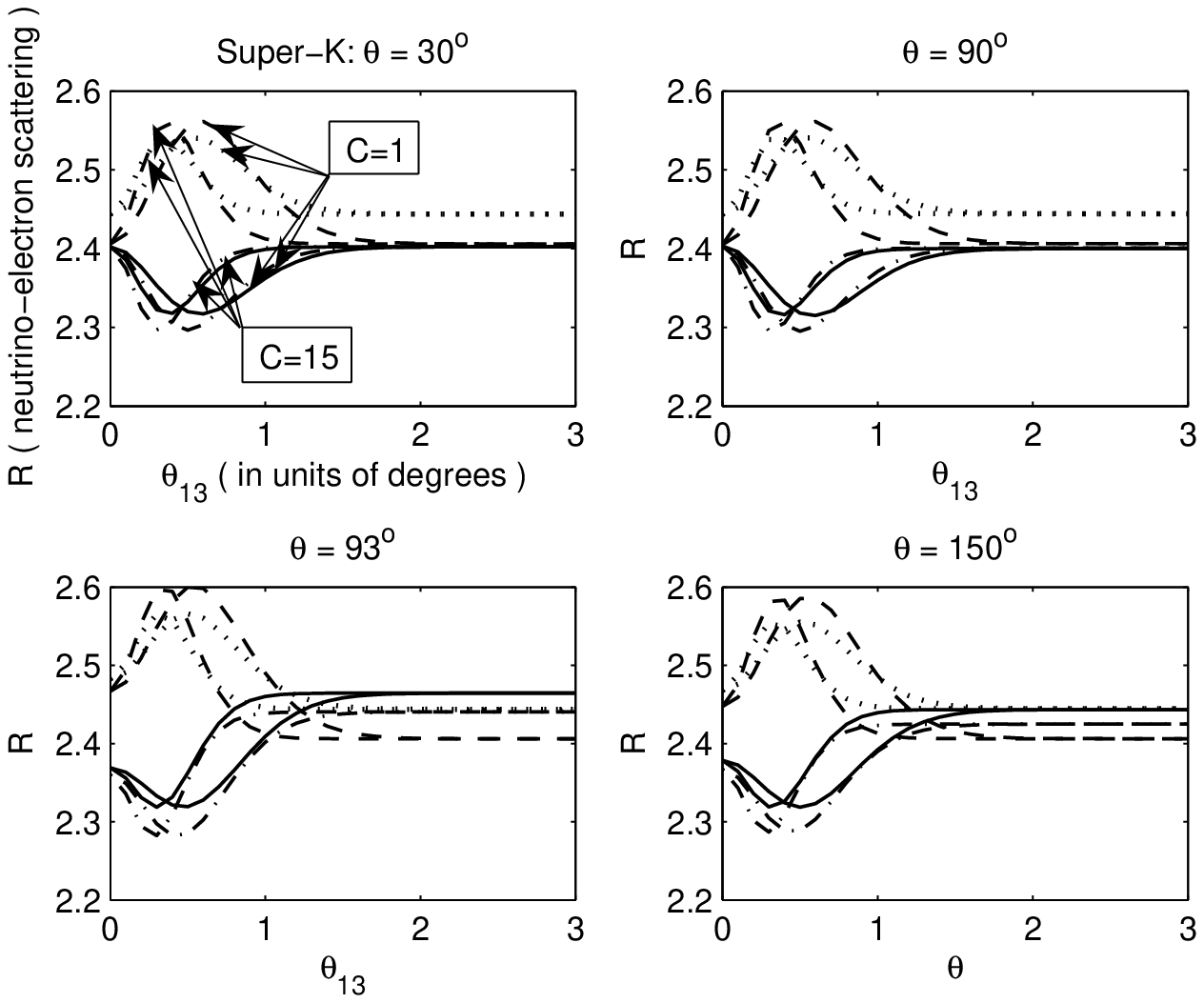}\\
\centerline{Fig. 9}
\end{figure}

\begin{figure}
\includegraphics[width=0.8\textwidth]{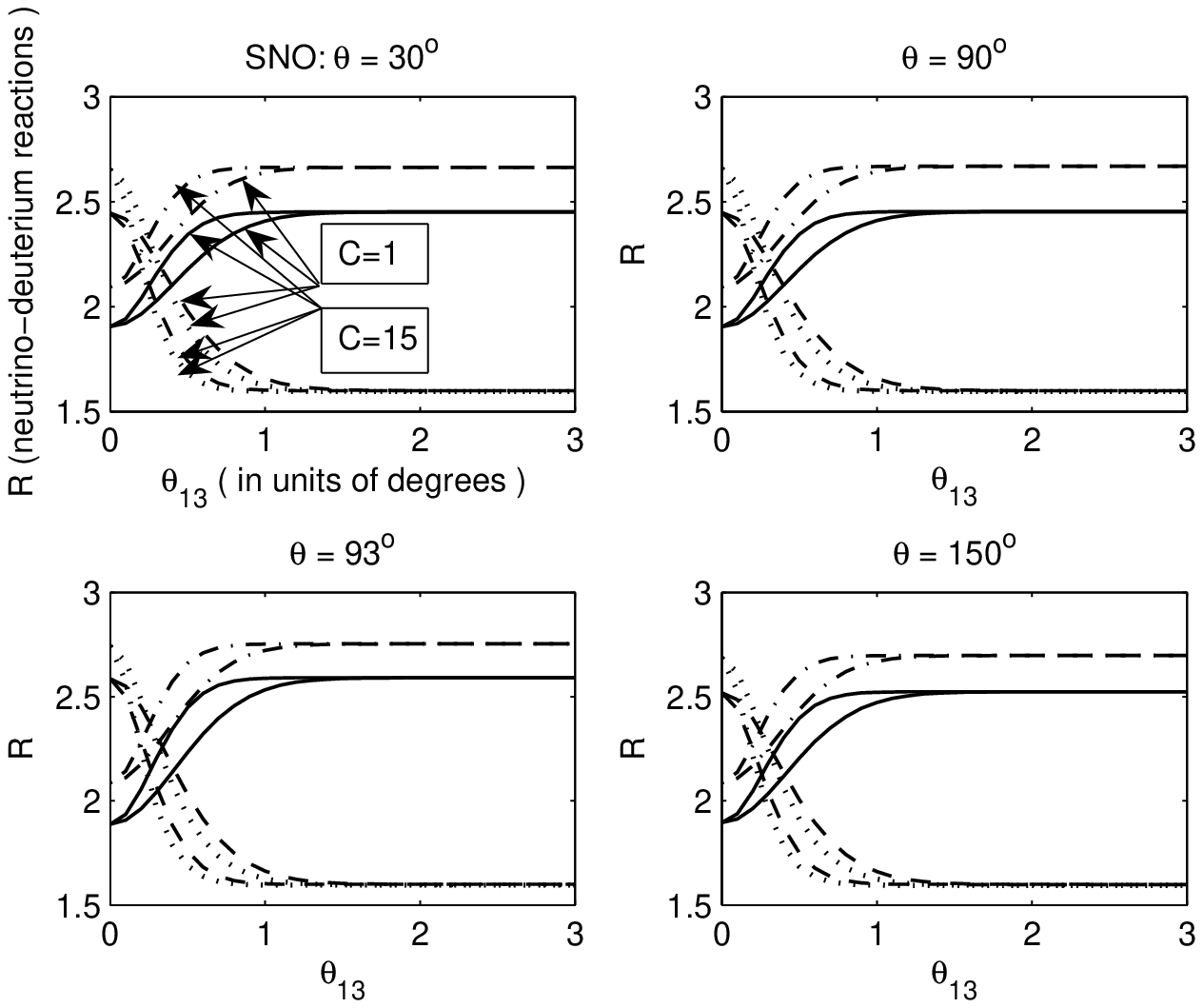}\\
\centerline{Fig. 10}
\end{figure}

\begin{figure}
\includegraphics[width=1\textwidth]{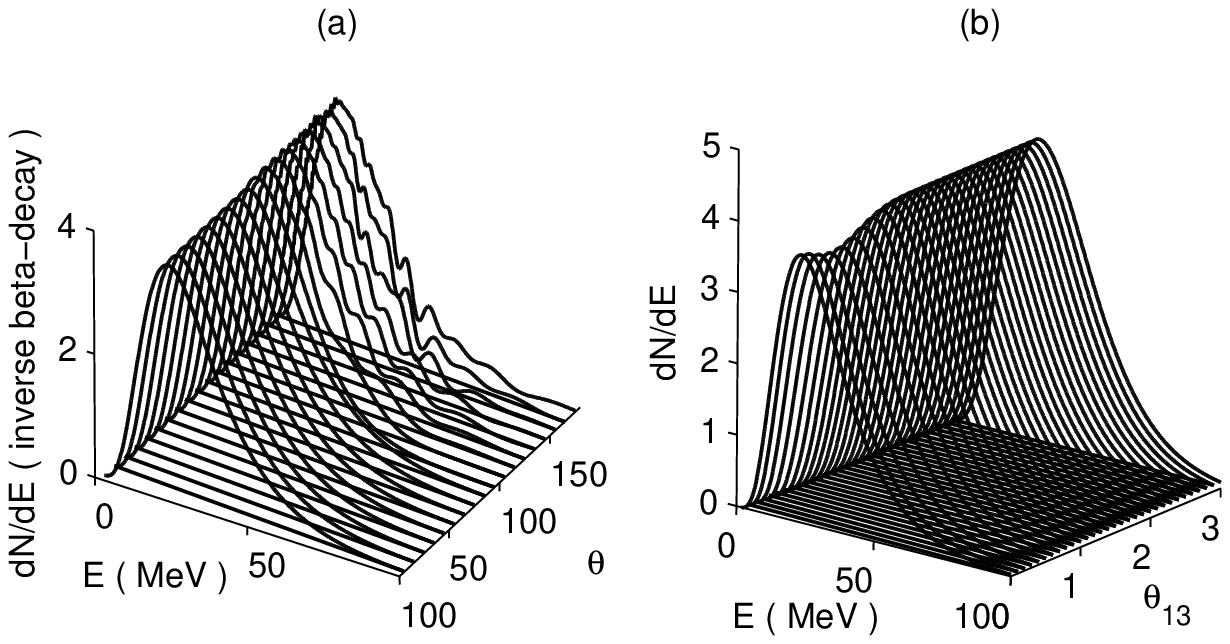}\\
\centerline{Fig. 11}
\end{figure}

\begin{table}[htb]
\begin{ruledtabular}
\caption{The realistic matter density inside the Earth where $r$
is the distance to the center of the Earth and $R=r(km)/6371$
\cite{Earth}.}
\begin{tabular}{rr}
$r(km)$ & $\rho$($10^3kg\cdot m^{-3}$)  \\
 \hline\hline
 $0-1221.5$ & $13.0885-8.8381R^2$\\
 $1221.5-3480.0$ & $12.5815-1.2638R-3.6426R^2-5.5281R^3$\\
 $3480.0-5701.0$ & $7.9565-6.4761R+5.5283R^2-3.0807R^3$\\
 $5701.0-5771.0$ & $5.3197-1.4836R$\\
 $5771.0-5971.0$ & $11.2494-8.0298R$\\
 $5971.0-6151.0$ & $7.1089-3.8045R$\\
 $6151.0-6346.6$ & $2.6910+0.6924R$\\
 $6346.6-6356.0$ & $2.900$\\
 $6356.0-6368.0$ & $2.600$\\
 $6368.0-6371.0$ & $1.020$\\
\end{tabular}
\end{ruledtabular}
\end{table}

\begin{table}[!htb]
\begin{ruledtabular}
\caption{Summary of the realistic Earth matter effects. $N$ ($I$)
represents the normal (inverted) hierarchy, $'1'$ $('0')$
represents $P_H=1$ $(0)$. The numbers in the columns 'Incipient'
and 'Min' are the event numbers when the SN neutrino Earch
incident angle is zero and is the angle in the column 'Angle',
respectively. The column 'Angle' gives the angles at which the
event numbers are minimum and the Earth matter effects are the
strongest. The column 'Ratio' gives the percentages of the Earth
matter effects. } \vspace{0.1cm}
\begin{tabular}{l|lllllll}
 Conditions & Hierarchy($P_H$) & Reaction  & Incipient & Min & Angle & Ratio \\
 \hline\hline
 {} & $N(1)$ & $\bar{\nu}_ep$  & 122.28 & 113.93
 & $94^\circ$ & 6.82\% \\
 {} & {} & $\nu e^-$  & 5.13 & 5.03 & $93^\circ$ &
 2.03\%\\
 {} & {} & $^{12}C$  & 36.70 & 35.98 & $93^\circ$ &
 1.97\%\\
 $T_{\nu_e}=3.5MeV$ & $I(1)$ & $\bar{\nu}_ep$  & 122.28 & 113.93
 & $94^\circ$ & 6.82\% \\
  $T_{\bar{\nu}_e}=5MeV$ & {} & $\nu e^-$  & 5.20 & 5.18 & $93^\circ$ &
 0.36\%\\
  $T_{\nu_x}=8MeV$ & {} & $^{12}C$ & 39.12 & 38.73 & $93^\circ$ &
 0.99\%\\
  $\eta_{\nu_e}=0$ & $N(0)$ & $\bar{\nu}_ep$ & 122.28 & 113.93
 & $94^\circ$ & 6.82\% \\
 $\eta_{\bar{\nu}_e}=0$ & {} & $\nu e^-$ & 5.13 & 5.11 & $93^\circ$ & 0.36\%\\
 $\eta_{\nu_x}=0$ & {} & $^{12}C$  & 39.18 & 38.79 & $93^\circ$ &
 0.99\%\\
 {} & $I(0)$ & $\bar{\nu}_ep$ & 171.72 & {} & {} & {}\\
 {} & {} & $\nu e^-$ & 5.20 & {} & {} & {}\\
 {} & {} & $^{12}C$ & 45.53 & {} & {} &
 {}\\
\hline
 {} & $N(1)$ & $\bar{\nu}_ep$ & 163.14 & 151.75 & $95^\circ$ & 6.98\%\\
 {} & {} & $\nu e^-$  & 5.13 & 5.00 & $94^\circ$ &
 2.48\%\\
 {} & {} & $^{12}C$ & 28.00 & 27.20 & $93^\circ$ & 2.86\%\\
 $T_{\nu_e}=4MeV$& $I(1)$ & $\bar{\nu}_ep$ & 163.14 & 151.75 & $95^\circ$ & 6.98\%\\
 $T_{\bar{\nu}_e}=6MeV$ & {} & $\nu e^-$  & 5.13 & 5.11 & $94^\circ$ &
 0.43\%\\
 $T_{\nu_x}=9MeV$ & {} & $^{12}C$ & 29.93 & 29.50 & $93^\circ$ & 1.43\%\\
 $\eta_{\nu_e}=5$ & $N(0)$ & $\bar{\nu}_ep$ & 163.14 & 151.75 & $95^\circ$ & 6.98\%\\
 $\eta_{\bar{\nu}_e}=2.5$ & {} & $\nu e^-$ & 5.13 & 5.11 & $94^\circ$ & 0.43\%\\
 $\eta_{\nu_x}=2$ & {} & $^{12}C$ & 29.96 & 29.53 & $93^\circ$ & 1.43\%\\
 {} & $I(0)$ & $\bar{\nu}_ep$ & 213.24 & {} & {} & {} \\
 {} & {} & $\nu e^-$ & 5.13 & {} & {} & {}\\
 {} & {} & $^{12}C$ & 35.16 & {} & {} & {}\\
 \hline
 {} & $N(1)$ & $\bar{\nu}_ep$ & 122.79 & 118.26 & $94^\circ$ & 3.69\%\\
 {} & {} & $\nu e^-$  & 5.13 & 5.05 & $93^\circ$ &
 1.59\%\\
 {}& {} & $^{12}C$ & 40.58 & 39.93 & $93^\circ$ & 1.62\%\\
 $T_{\nu_e}=3MeV$ & $I(1)$ & $\bar{\nu}_ep$ & 122.79 & 118.26 & $94^\circ$ & 3.69\%\\
 $T_{\bar{\nu}_e}=5MeV$ & {} & $\nu e^-$  & 5.17 & 5.16 & $93^\circ$ &
 0.21\%\\
 $T_{\nu_x}=7MeV$ & {} & $^{12}C$ & 43.24 & 42.89 & $92^\circ$ & 0.81\%\\
 $\eta_{\nu_e}=3$ & $N(0)$ & $\bar{\nu}_ep$ & 122.79 & 118.26 & $94^\circ$ & 3.69\%\\
 $\eta_{\bar{\nu}_e}=2$ & {} & $\nu e^-$ & 5.13 & 5.12 & $93^\circ$ & 0.21\%\\
 $\eta_{\nu_x}=0$  & {} & $^{12}C$ & 43.45 & 43.10 & $92^\circ$ & 0.80\%\\
 {} & $I(0)$ & $\bar{\nu}_ep$ & 148.53 & {} & {} & {}\\
 {} & {} & $\nu e^-$ & 5.17 & {} & {} & {}\\
 {}& {} & $^{12}C$ & 51.38 & {} &  {} &  {}\\
 \end{tabular}
 \end{ruledtabular}
 \end{table}
\end{document}